\documentclass[twocolumn]{revtex4}
\usepackage[dvips]{graphicx}
\usepackage{float, color,array, multirow}

\begin{document}

\title{Effects of spin-lattice coupling and a magnetic field in classical Heisenberg antiferromagnets on the breathing pyrochlore lattice}

\author{Kazushi Aoyama$^1$, Masaki Gen$^2$, and Hikaru Kawamura$^3$}

\date{\today}

\affiliation{$^1$Department of Earth and Space Science, Graduate School of Science, Osaka University, Osaka 560-0043, Japan \\
$^2$Department of Advanced Materials Science, University of Tokyo, Kashiwa 277-8561, Japan \\
$^3$Molecular Photoscience Research Center, Kobe University, Kobe 657-8501, Japan
}

\begin{abstract}
We theoretically investigate spin-lattice coupling (SLC) effects on the in-field ordering properties of classical Heisenberg antiferromagnets on the breathing pyrochlore lattice. Here, we use the two possible simplified models describing the effect of local lattice distortions on the spin ordering via the SLC, the bond-phonon and site-phonon models. It is found by means of Monte Carlo simulations that in both models, the $\frac{1}{2}$ plateau shows up in the magnetization curve being relatively robust against the breathing bond-alternation, although magnetic long-range orders (LRO's) are realized only in the site-phonon model. In the bond-phonon model, additional further neighbor interactions are necessary to induce a magnetic LRO. In the site-phonon model, it is also found that in addition to the low-field, middle-field $\frac{1}{2}$-plateau, and high-field phases appearing on both the uniform and breathing pyrochlore lattices, various types of unconventional phases which can be viewed as LRO's in units of tetrahedron are induced by the breathing bond-alternation just below the $\frac{1}{2}$ plateau and the saturation field. The occurrence of these tetrahedron-based orders could be attributed to the nature characteristic of the breathing pyrochlore lattice, i.e., the existence of the nonequivalent small and large tetrahedra. Experimental implications of our result are also discussed.  
\end{abstract}

\maketitle
\section{introduction}
Frustrated magnets often exhibit a magneto-structural transition in which the system releases the magnetic frustration by spontaneously distorting the underlying lattice, resulting in a magnetic long-range order (LRO). Such a spin-lattice-coupled ordering can commonly be seen in a series of spinel chromium oxides $A$Cr$_2$O$_4$ ($A$=Hg, Cd, Zn, Mg) \cite{ZnCrO_Lee_00, CdCrO_Chung_05, HgCrO_Ueda_06, MgCrO_Ortega_08}, where the magnetic Cr$^{3+}$ sites form the pyrochlore lattice, a three-dimensional network consisting of corner-sharing tetrahedra. The characteristic properties of these compounds are a first-order simultaneous magnetic and structural transition at zero field \cite{ZnCrO_Lee_00, CdCrO_Chung_05, HgCrO_Ueda_06, MgCrO_Ortega_08} and a field-induced $\frac{1}{2}$-magnetization-plateau phase \cite{HgCrO_Ueda_06,CdCrO_Kojima_08,CdCrO_Miyata_13,ZnCrO_Miyata_jpsj_11,ZnCrO_Miyata_prl_11,ZnCrO_Miyata_jpsj_12,HgCrO_Nakamura_jpsj_14,MgCrO_Miyata_jpsj_14} which is considered to originate from the spin-lattice coupling (SLC). Similar zero-field and in-field properties have been observed also in LiInCr$_4$O$_8$ which belongs to a new class of chromium oxides  hosting the so-called breathing pyrochlore lattice, an alternation array of small and large tetrahedra \cite{BrPyro_Okamoto_13,BrPyro_Tanaka_14,BrPyro_Nilsen_15,BrPyro_Saha_16,BrPyro_Lee_16,BrPyro_Hdep_Okamoto_17,BrPyro_Hdep_Gen_19}. In this paper, we theoretically investigate effects of the magnetic field on the spin-lattice-coupled ordering in the breathing-pyrochlore antiferromagnets, based on the two possible simplified models describing the SLC, the bond-phonon \cite{Bond_Penc_04,Bond_Motome_06,Bond_Shannon_10} and site-phonon \cite{Site_Jia_05,Site_Bergman_06,Site_Wang_08,Site_AK_16,Site_AK_19} models which will be explained below.      

In the chromium oxides, Hund-coupled three $3d$ electrons at each Cr$^{3+}$ site occupy the 3-fold $t_{2g}$ level, constituting a localized $S=3/2$ spin with their orbital degrees of freedom off. As the magnetic anisotropy is negligible, the classical Heisenberg model should provide a reasonable modeling. In the nearest neighbor (NN) antiferromagnetic classical Heisenberg model, it is theoretically well established that any magnetic LRO does not occur at any finite temperature due to a massive ground-state degeneracy \cite{Reimers_MC_92,Moessner-Chalker_prl,Moessner-Chalker_prb}, but in the $A$Cr$_2$O$_4$ family, the degeneracy is lifted, via SLC, by lattice distortions which lower the lattice symmetry from cubic to tetragonal or orthorhombic, leading to antiferromagnetic LRO's \cite{ZnCrO_Lee_00, CdCrO_Chung_05, HgCrO_Ueda_06, MgCrO_Ortega_08}. Besides, in a magnetic field, $A$Cr$_2$O$_4$ commonly show the $\frac{1}{2}$-magnetization plateau \cite{HgCrO_Ueda_06,CdCrO_Kojima_08,CdCrO_Miyata_13,ZnCrO_Miyata_jpsj_11,ZnCrO_Miyata_prl_11,ZnCrO_Miyata_jpsj_12,HgCrO_Nakamura_jpsj_14,MgCrO_Miyata_jpsj_14}, pointing to a robust 3-up and 1-down collinear spin configuration on each tetrahedron. Such a collinear state is considered to be stabilized by the biquadratic interaction of the from $-({\bf S}_i \cdot {\bf S}_j)^2$ originating from the SLC \cite{Bond_Penc_04,Bond_Motome_06,Bond_Shannon_10}. In this $\frac{1}{2}$-plateau phase, the crystal structure changes to cubic in the Hg and Cd compounds \cite{HgCrO_Matsuda_07,CdCrO_Inami_06,CdCrO_Matsuda_prl_10}. These zero-field and in-field experimental results suggest that SLC is essential in the uniform pyrochlore antiferromagnets $A$Cr$_2$O$_4$. 

\begin{figure*}[t]
\begin{center}
\includegraphics[scale=0.80]{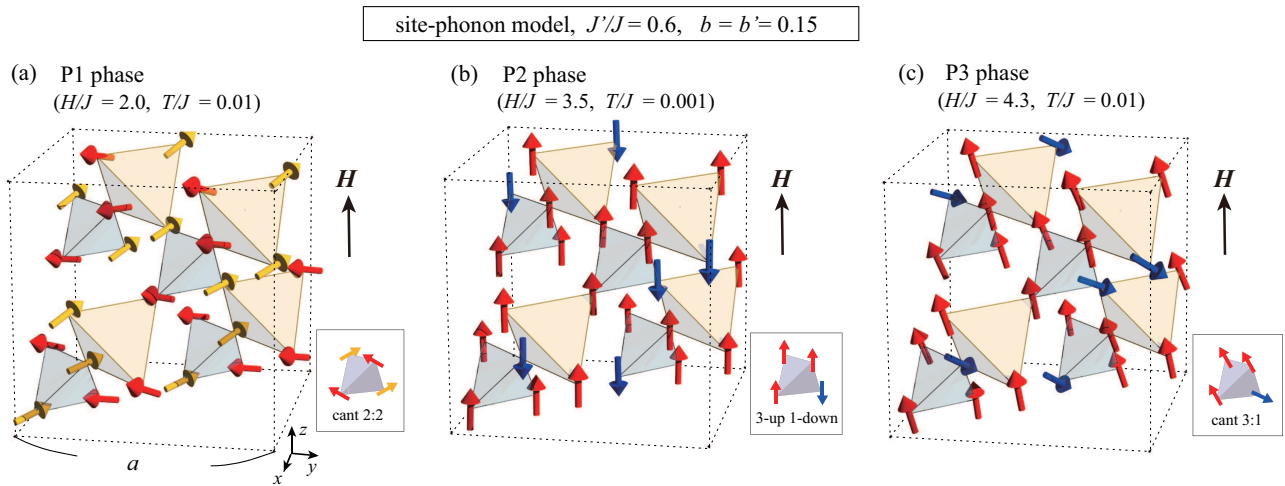}
\caption{Real-space spin structures of the low-field P1, middle-field P2, and high-field P3 phases obtained in the MC simulations for the site-phonon model (\ref{eq:Hamiltonian_SP}) with $J'/J=0.6$ and $b=b'=0.15$ at (a) $H/J=2.0$, (b) $H/J=3.5$, and (c) $H/J=4.3$. Red and yellow (blue) arrows represent spins pointing upward (downward) along the applied magnetic field $H$. In (a), the in-plane components of the red and yellow spins are antiparallel to each other. In the P1, P2, and P3 phases, local spin configurations on each tetrahedron are the cant 2:2, 3-up 1-down, and cant3:1, respectively (see the inset of each figure). These three 16-sublattice states are realized on both the {\it uniform} and {\it breathing} pyrochlore lattices. Concerning the spin structures of other phases unique to the breathing system (SP1, SP2, SP2', SP3, and SP4 phases in Fig. \ref{fig:HT_siteall}), see Fig. \ref{fig:snap_site020J06} in the main text and Figs. \ref{fig:snap_site020J02H0180} and \ref{fig:snap_site020J02H0350-0370} in Appendix C.  \label{fig:snap_site015J06}}
\end{center}
\end{figure*}

Breathing-pyrochlore magnets also provide examples of the spin-lattice-coupled ordering. Among so-far reported several compounds \cite{BrPyro_Saha_17,BrPyro_doped_Okamoto_15,BrPyro_doped_Wang_17,BrPyro_doped_Wawrzynczak_17,BrPyro_Sulfides_Okamoto_18,BrPyro_Sulfides_Pokharel_18,BrPyro_Hdep_Gen_20,BrPyro_Sulfides_Kanematsu_20,BrPyro_Sulfides_Pokharel_20,qBrPyro_Kimura_14,qBrPyro_Haku_prb16,qBrPyro_Haku_jpsj16,qBrPyro_Rau_16,qBrPyro_Rau_18}, the chromium oxides Li(Ga, In)Cr$_4$O$_8$ \cite{BrPyro_Okamoto_13,BrPyro_Tanaka_14,BrPyro_Nilsen_15,BrPyro_Saha_16,BrPyro_Lee_16,BrPyro_Hdep_Okamoto_17,BrPyro_Hdep_Gen_19} exhibit magnetic properties similar to those of $A$Cr$_2$O$_4$.   
In Li(Ga, In)Cr$_4$O$_8$, the NN interactions on small and large tetrahedra, $J$ and $J'$, are antiferromagnetic with different strength, where the ratio $J'/J$ is estimated from the bond-length difference to be $J'/J \sim 0.1$ and $0.6$ for the In and Ga compounds, respectively \cite{BrPyro_Okamoto_13}. In these compounds, the massive ground-state degeneracy, which is still present at the level of the NN model with different antiferromagnetic $J$ and $J'$ \cite{BrPyro_NNmodel_Benton_15}, is lifted by distorting the lattice from cubic to tetragonal, although in contrast to the uniform case of $A$Cr$_2$O$_4$, the structural transition slightly preempts the magnetic one \cite{BrPyro_Tanaka_14,BrPyro_Nilsen_15,BrPyro_Saha_16,BrPyro_Lee_16}. We note that according to Refs. \cite{BrPyro_Nilsen_15,BrPyro_Saha_16}, the structural transition in these breathing systems is incomplete and the low-temperature ordered phase is a coexistence of the original cubic and emergent tetragonal crystal domains. In addition to this zero-field property, recent high-field measurements on LiInCr$_4$O$_8$ show the occurrence of the $\frac{1}{2}$-magnetization plateau \cite{BrPyro_Hdep_Okamoto_17,BrPyro_Hdep_Gen_19}, which suggests that the SLC is also important in the breathing-pyrochlore chromium oxides. Interestingly, in the chromium sulfide CuInCr$_4$S$_8$ with antiferromagnetic $J$ and ferromagnetic $J'$, the $\frac{1}{2}$ plateau has also been observed and the effect of the SLC has been pointed out \cite{BrPyro_Hdep_Gen_20}.  
In this paper, bearing the chromium oxides in our mind, we will theoretically investigate the effects of both the SLC and the external magnetic field on the spin ordering in the breathing pyrochlore antiferromagnets with antiferromagnetic $J$ and $J'$. 

Theories of the SLC in pyrochlore antiferromagnets could be classified into two. One is a phenomenological theory based on the group theoretical classification of the lattice distortion \cite{SLC_Yamashita_00,SLC_Tchernyshyov_prl_02,SLC_Tchernyshyov_prb_02}, and the other is a microscopic theory taking account of the effect of {\it local} lattice distortions which have been modeled alternatively by bond phonons or site phonons. In this work, we take the latter microscopic approach. In the bond-phonon model which was first introduced by Penc {\it et al}. \cite{Bond_Penc_04,Bond_Motome_06,Bond_Shannon_10} and has conventionally been used to describe the SLC effect, each bond is assumed to vibrate independently, whereas in the site-phonon model \cite{Site_Jia_05,Site_Bergman_06,Site_Wang_08,Site_AK_16,Site_AK_19}, the lattice vibration is modeled by the Einstein phonon, i.e., each site is assumed to vibrate independently. The bond phonon involves only two spins ${\bf S}_i$ and ${\bf S}_j$ residing on both ends of each bond, giving rise to the effective spin interaction of the biquadratic form $-({\bf S}_i \cdot {\bf S}_j)^2$. The site phonon, on the other hand, additionally involves inter-bond spins, so that it mediates effective further neighbor interactions in addition to the biquadratic one. 

In both the uniform and breathing pyrochlore antiferromagnets at zero field, the {\it site-phonon} system undergoes a first-order transition into a collinear magnetic LRO which is characterized by $(1,1,0)$-type [$(\frac{1}{2},\frac{1}{2},\frac{1}{2})$-type] magnetic Bragg peaks for weak (strong) SLC \cite{Site_AK_16, Site_AK_19}. The $(1,1,0)$ state realized in the weak SLC regime, which corresponds to the antiferromagnetic order with the lattice distortion of the $E_u$ phonon in Ref. \cite{SLC_Tchernyshyov_prb_02}, has the same spin structure as that observed in the tetragonal crystal domains of the breathing pyrochlore antiferromagnets Li(Ga, In)Cr$_4$O$_8$ \cite{BrPyro_Nilsen_15,BrPyro_Saha_16}. Furthermore, although the $(1,1,0)$ state itself does not seem to be reported in the uniform pyrochlore antiferromagnets $A$Cr$_2$O$_4$ \cite{CdCrO_Chung_05,ZnCrO_Lee_08,HgCrO_Matsuda_07,MgCrO_Ortega_08}, it has been shown that a $(1,1,0)$-like state slightly modified by the Dzyaloshinskii-Moriya (DM) interaction is consistent with the Neel state of CdCr$_2$O$_4$ \cite{SLC_Chern_06}. Thus, the site-phonon model should capture an essential feature of the SLC in the chromium oxides. In the bond-phonon model, on the other hand, any magnetic LRO does not appear and only a spin nematic state is realized in the low-temperature ordered phase, so that to induce a magnetic LRO, additional further neighbor interactions need to be incorporated \cite{Bond_Shannon_10}. Indeed, in Refs. \cite{Bond_Motome_06,Bond_Shannon_10,ZnCrO_Miyata_jpsj_11,ZnCrO_Miyata_jpsj_12} taking the bond-phonon picture, the additional ferromagnetic third NN interaction is incorporated as the simplest example, although the obtained magnetic LRO at zero field is not consistent with the experimental result. 
In spite of the difference in the ordering properties at zero field, the two models share a common in-field feature at least on the uniform pyrochlore lattice. The biquadratic interaction common to both models favors the collinear spin state, stabilizing the $\frac{1}{2}$-magnetization-plateau phase \cite{Bond_Penc_04,Bond_Motome_06,Bond_Shannon_10,Site_Bergman_06}, which is consistent with the experimental observation in $A$Cr$_2$O$_4$ \cite{HgCrO_Ueda_06,CdCrO_Kojima_08,CdCrO_Miyata_13,ZnCrO_Miyata_jpsj_11,ZnCrO_Miyata_prl_11,ZnCrO_Miyata_jpsj_12,HgCrO_Nakamura_jpsj_14,MgCrO_Miyata_jpsj_14}. Furthermore, it has been shown by Bergman {\it et al}. \cite{Site_Bergman_06} that the site phonon stabilizes the same magnetic structure as that observed in the $\frac{1}{2}$-plateau phase of HgCr$_2$O$_4$ and CdCr$_2$O$_4$ \cite{HgCrO_Matsuda_07,CdCrO_Matsuda_prl_10}. Then, the natural question is how the magnetization processes in the two models behave in the breathing case. In this work, we examine the in-field properties of both the bond-phonon and site-phonon models on the breathing pyrochlore lattice, focusing on the weak SLC regime which should be relevant to the existing materials Li(Ga, In)Cr$_4$O$_8$. 
  
It will be shown by means of Monte Carlo (MC) simulations that in both the two models, the $\frac{1}{2}$ plateau is robust against the breathing bond-alternation, but that whether the spin-lattice-coupled orderings are affected or not depends on the model. In the bond-phonon model, the ordering properties are not altered by the breathing bond-alternation: any magnetic LRO does not appear, and two types of quadruple orders as well as a spin-liquid-plateau phase appear as in the uniform case \cite{Bond_Shannon_10}. In the site-phonon model, on the other hand, three magnetically long-range-ordered states, the low-field, middle-field $\frac{1}{2}$-plateau, and high-field phases (see P1, P2, and P3 phases in Figs. \ref{fig:snap_site015J06}, \ref{fig:GS}, and \ref{fig:HT_siteall}), appear in the wide range of the parameter space on both the {\it uniform} and {\it breathing} pyrochlore lattices. Local spin configurations on each tetrahedron in the three phases are the so-called cant 2:2, 3-up 1-down, and cant 3:1, respectively. In addition to the three, the breathing bond-alternation can induce unconventional phases just below the $\frac{1}{2}$-plateau phase and the saturation field (see SP1, SP2, SP2', SP3, and SP4 phases in Fig. \ref{fig:HT_siteall}). In contrast to the basic three phases shown in Fig. \ref{fig:snap_site015J06} where the spin configuration on each tetrahedron is equivalent to one another, in the SP1, SP2, SP2', SP3, and SP4 phases (for their real-space structures, see Figs. \ref{fig:snap_site020J06}, \ref{fig:snap_site020J02H0180}, and \ref{fig:snap_site020J02H0350-0370}), the spin configuration on each tetrahedron is not equivalent any more, resulting in tetrahedron-based LRO's, which could be attributed to the nature characteristic of the breathing pyrochlore lattice, i.e., the existence of the nonequivalent small and large tetrahedra.  

The outline of this paper is as follows: In Sec. II, we introduce the models taking account of the local lattice distortions, i.e., the bond-phonon and site-phonon models, and derive their effective spin Hamiltonians. Physical quantities relevant to the present system and numerical methods are explained in Sec. III. This is followed by Secs. IV and V in which the ordering properties of the bond-phonon and site-phonon models are respectively, discussed. We end the paper with summary and discussion in Sec. VI. In the Appendix, details of the numerical method and real-space structures of several magnetic LRO's are explained.

\section{model}
In this section, we derive the effective spin Hamiltonian describing the SLC in the presence of the breathing bond-alternation. Throughout this paper, the NN sites denote the neighboring sites connected by a bond independent of its length. 
Suppose that the displacement vector at each site $i$ from its regular position ${\bf r}^0_i$ on the lattice is denoted by ${\bf u}_i$, 
a minimal microscopic Hamiltonian could be written as
\begin{equation}\label{eq:original_H}
{\cal H} = \sum_{\langle i,j \rangle } J_{\rm ex}\big(|{\bf r}^0_{ij} + {\bf u}_i-{\bf u}_j|\big){\bf S}_i \cdot {\bf S}_j -H\sum_i S_i^z+ {\cal H}_{\rm L},
\end{equation}
where ${\bf S}_i$ is the classical Heisenberg spin at the site $i$, ${\bf r}^0_{ij} \equiv {\bf r}^0_i-{\bf r}^0_j$, 
$J_{\rm ex}$ is the exchange interaction which is, for simplicity, assumed to depend only on the distance between the two spins, the summation $\langle i,j \rangle$ is taken over all the NN sites, $H$ is a magnetic field applied in the $z$-direction in the spin space, and ${\cal H}_{\rm L}$ denotes the elastic energy. 
As the displacement is usually small, i.e., $|{\bf u}_i|/|{\bf r}^0_i| \ll 1$, we can expand the exchange interaction with respect to the displacement as follows:
{\begin{eqnarray}\label{eq:expansion}
J_{\rm ex}\big(|{\bf r}^0_{ij} + {\bf u}_i-{\bf u}_j| \big) &=& J_{\rm ex}\big(|{\bf r}^0_{ij}|\big) + \frac{d J_{\rm ex}}{dr}\Big|_{r=|{\bf r}^0_{ij}|} \, {\bf e}_{ij} \cdot ({\bf u}_i-{\bf u}_j ) \nonumber\\
&+& O(\big[{\bf e}_{ij} \cdot ({\bf u}_i-{\bf u}_j )\big]^2)
\end{eqnarray}
where ${\bf e}_{ij} \equiv {\bf r}^0_{ij}/|{\bf r}^0_{ij}|$ is the unit vector connecting NN sites $i$ and $j$. Hereafter, we take the leading-order correction of the order of $O({\bf e}_{ij} \cdot ({\bf u}_i-{\bf u}_j ))$ in Eq. (\ref{eq:expansion}) into account.

\begin{figure}[t]
\begin{center}
\includegraphics[scale=0.3]{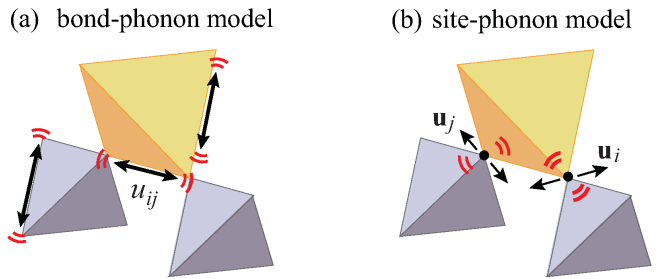}
\caption{Two possible minimal models describing the local lattice distortions. (a) Bond-phonon model taking account of the independent bond-length change $u_{ij}$. (b) Site-phonon model taking account of the independent site displacement ${\bf u}_i$.\label{fig:model}}
\end{center}
\end{figure}

Concerning the elastic energy ${\cal H}_{\rm L}$, although in reality neighboring ${\bf u}_i$'s should be correlated to each other in the form of dispersive phonon modes, we use here the local phonon models, the bond-phonon and site-phonon models, because they are possible minimum tractable models describing phonon-mediated spin interactions. In the bond-phonon and site-phonon models, variables describing the lattice degrees of freedom are the length change of each bond $u_{ij}\equiv{\bf e}_{ij}\cdot\big({\bf u}_i-{\bf u}_j \big)$ and the displacement at each site ${\bf u}_i$, respectively (see Fig. \ref{fig:model}), and the elastic energy ${\cal H}_{\rm L}$ in each model is given by 
\begin{equation}\label{eq:lattice_H}
{\cal H}_{\rm L} =  \left \{\begin{array}{l}
\displaystyle{ \frac{c_{\rm BP}}{2}\sum_{\langle i,j \rangle} u_{ij}^2 } \qquad (\mbox{bond-phonon model}), \nonumber\\
\displaystyle{ \frac{c_{\rm SP}}{2}\sum_i |{\bf u}_i|^2 } \qquad (\mbox{site-phonon model}),
\end{array} \right . 
\end{equation}
with elastic constants $c_{\rm BP}$ and $c_{\rm SP}$ \cite{Bond_Penc_04,Site_Bergman_06}. Note that compared with the conventional Deby model, the bond-phonon model assumes that variables describing lattice degrees of freedom are not ${\bf u}_i$'s but $u_{ij}$'s with the elastic energy ${\cal H}_{\rm L}$ being the same as that of the Deby phonon, whereas the site-phonon (Einstein-phonon) model assumes that the elastic energy is a local one with the lattice degrees of freedom ${\bf u}_i$'s being unchanged.
Substituting Eq. (\ref{eq:expansion}) into Eq. (\ref{eq:original_H}) and integrating out the lattice degrees of freedom $u_{ij}$ for the bond-phonon model and ${\bf u}_i$ for the site-phonon model, we obtain the bond-phonon spin Hamiltonian ${\cal H}_{\rm BP}$ as
\begin{eqnarray}\label{eq:Hamiltonian_BP}
{\cal H}_{\rm BP} &=& J \, \sum_{\langle i,j \rangle_S } {\bf S}_i \cdot {\bf S}_j + J' \, \sum_{\langle i,j \rangle_L } {\bf S}_i \cdot {\bf S}_j -H\sum_i S_i^z \nonumber\\
&-& J \, b \, \sum_{\langle i,j \rangle_S } \big( {\bf S}_i \cdot {\bf S}_j \big)^2 - J' \, b' \,\sum_{\langle i,j \rangle_L } \big( {\bf S}_i \cdot {\bf S}_j \big)^2 , 
\end{eqnarray}
and the site-phonon spin Hamiltonian ${\cal H}_{\rm SP}$ as
\begin{eqnarray}\label{eq:Hamiltonian_SP}
{\cal H}_{\rm SP} &=& J \, \sum_{\langle i,j \rangle_S } {\bf S}_i \cdot {\bf S}_j + J' \, \sum_{\langle i,j \rangle_L } {\bf S}_i \cdot {\bf S}_j -H\sum_i S_i^z\\
&-& J \, b \, \sum_{\langle i,j \rangle_S } \big( {\bf S}_i \cdot {\bf S}_j \big)^2 - J' \, b' \,\sum_{\langle i,j \rangle_L } \big( {\bf S}_i \cdot {\bf S}_j \big)^2 \nonumber\\
&-& \sum_i \Big\{ \frac{Jb}{4} \sum_{j\neq k \in N_S(i)}+\frac{J'b'}{4}\sum_{j\neq k \in N_L(i)}\Big\} \big( {\bf S}_i \cdot {\bf S}_j \big)\big( {\bf S}_i \cdot {\bf S}_k \big) \nonumber\\
&-& \sqrt{J \, J' \,b \, b'}\sum_i \sum_{j \in N_S(i) } \sum_{k \in N_L(i) } {\bf e}_{ij} \cdot {\bf e}_{ik} \, \big( {\bf S}_i \cdot {\bf S}_j \big)\big( {\bf S}_i \cdot {\bf S}_k \big),  \nonumber
\end{eqnarray}
where two kinds of NN exchange interactions $J\equiv J_{\rm ex}\big(|{\bf r}^0_{ij}|_{\rm Small}\big)$ and $J'\equiv J_{\rm ex}\big(|{\bf r}^0_{ij}| _{\rm Large}\big)$, NN sites $N_{S}(i)$ and $N_{L}(i)$, and the summations $\langle i,j \rangle _{S}$ and $\langle i,j \rangle _{L}$ are defined only on the small and large tetrahedra, respectively. The degree of the breathing lattice-distortion is quantified by the ratio $0<J'/J \leq 1$, and the dimensionless parameters
\begin{eqnarray}\label{eq:b-def}
b &=& \frac{1}{cJ}\Big[ \frac{d J_{\rm ex}}{dr}\big|_{r=|{\bf r}^0_{ij}|_{\rm Small}} \Big]^2 \nonumber\\
b' &=& \frac{1}{cJ'}\Big[ \frac{d J_{\rm ex}}{dr}\big|_{r=|{\bf r}^0_{ij}|_{\rm Large}} \Big]^2 
\end{eqnarray}
measure the strength of the SLC for small and large tetrahedra, respectively, where $c=2c_{\rm BP}$ in the bond-phonon model and $c=c_{\rm SP}$ in the site-phonon model. We take $J, \, J'>0$ and $d J_{\rm ex}/dr < 0$, so that $b, \, b' >0$. In the uniform case of $J'/J=1$, there is, of course, no distinction between the small and large tetrahedra, so that there is only one SLC parameter, i.e., $b=b'$.

The feature common to the two models is the existence of the biquadratic interaction of the form $-({\bf S}_i \cdot {\bf S}_j)^2$. Since the overall sign of this interaction is always negative irrespective of the signs of $J$ and $J'$, the $-({\bf S}_i \cdot {\bf S}_j)^2$ term tends to align neighboring spins to be collinear and is known to be an origin of the spin nematic state. In contrast to the bond-phonon model (\ref{eq:Hamiltonian_BP}), the site-phonon model (\ref{eq:Hamiltonian_SP}) contains additional intra-tetrahedron interactions [the third line in Eq. (\ref{eq:Hamiltonian_SP})] and inter-tetrahedron ones [the fourth line in Eq. (\ref{eq:Hamiltonian_SP})]. Due to the inter-tetrahedron interactions involving further neighbor spins, a magnetic LRO becomes possible in the site-phonon model. By contrast, as we will demonstrate in Sec. IV, the bond-phonon model does not exhibit any magnetic LRO because of the absence of such effective further neighbor interactions.

\section{Relevant physical quantities and numerical method}
In this section, we first introduce physical quantities relevant to our systems described by the Hamiltonians (\ref{eq:Hamiltonian_BP}) and (\ref{eq:Hamiltonian_SP}), and then, explain the numerical method to calculate them. Throughout this paper, $N$ is a total number of spins and $\langle {\cal O} \rangle$ denotes the thermal average of a physical quantity ${\cal O}$.
\subsection{relevant physical quantities}
Since both the bond-phonon and site-phonon models have the biquadratic term $-({\bf S}_i \cdot {\bf S}_j)^2$ favoring collinear spin states, a key physical quantity is the spin collinearity which can be measured by   
\begin{equation}\label{eq:OP_nematic}
P = \frac{3}{2} \Big\langle \frac{1}{N^2}\sum_{i,j} \big( {\bf S}_i\cdot{\bf S}_j\big)^2 - \frac{1}{3} \Big\rangle. 
\end{equation}
As the magnetic field $H$ is applied in the $S^z$ direction, it is convenient to divide $P$ into the spin collinearity for the direction parallel to the field $P_{\parallel}$ and the ones for the perpendicular direction $P_{\perp 1}$ and $P_{\perp 2}$ as follows:
\begin{eqnarray}\label{eq:OP_nematic_div}
P &=& P_\parallel + P_{\perp 1} + P_{\perp 2}, \nonumber\\
P_\parallel  &=& \frac{3}{4}\Big\langle (Q^{3z^2-r^2})^2 \Big\rangle , \nonumber\\
P_{\perp 1}  &=& \frac{3}{4}\Big\langle |{\bf Q}_{\perp 1}|^2 \Big\rangle, \quad {\bf Q}_{\perp 1}=(Q^{xz},Q^{yz}), \nonumber\\
P_{\perp 2}  &=& \frac{3}{4}\Big\langle |{\bf Q}_{\perp 2}|^2 \Big\rangle, \quad {\bf Q}_{\perp 2}=(Q^{x^2-y^2}, Q^{xy}) .
\end{eqnarray}  
Here, we have introduced the rank-two tensor order parameters $Q^\alpha$ defined by $Q^\alpha = \frac{1}{N}\sum_{i} Q^\alpha_i$ with the local quadrupole moments \cite{Bond_Shannon_10}
\begin{eqnarray}
Q^{3z^2-r^2}_i &=& \frac{1}{\sqrt{3}}\big\{ 2(S_i^z)^2- (S_i^x)^2 - (S_i^y)^2 \big\}, \nonumber\\
Q^{x^2-y^2}_i &=&  (S_i^x)^2- (S_i^y)^2, \nonumber\\
Q^{xy}_i &=& 2S_i^x S_i^y, \nonumber\\
Q^{xz}_i &=& 2S_i^x S_i^z, \nonumber\\
Q^{yz}_i &=& 2S_i^y S_i^z .
\end{eqnarray}
For a spin-space rotation in the $S^xS^y$-plane by the angle $\phi$, ${\bf Q}_{\perp n}$ is rotated by the angle $n \phi$, so that $P_{\perp n}$ can be used as an order parameter to detect the
$n$-fold breaking of rotational symmetry in the $S^xS^y$ plane. When a magnetic (dipolar) LRO is absent but a quadruple LRO characterized by nonzero value of $P_{\perp 1}$ ($P_{\perp 2}$) exists, such a LRO is called the vector-multipole (nematic) order \cite{Bond_Shannon_10}.

Whether a magnetic (dipolar) LRO is present or not can be examined by measuring the spin structure factors  
\begin{eqnarray}\label{eq:F_S}
F_{S\parallel}({\bf q}) &=& \Big\langle \Big| \frac{1}{N} \sum_i  S^z_i \, e^{i{\bf q}\cdot{\bf r}^0_i}\Big|^2\Big\rangle, \nonumber\\
F_{S\perp}({\bf q}) &=& \Big\langle \sum_{\nu=x,y} \Big| \frac{1}{N} \sum_i  S^\nu_i \, e^{i{\bf q}\cdot{\bf r}^0_i}\Big|^2\Big\rangle. 
\end{eqnarray}
Noting that the magnetic field $H$ is applied in the $S^z$ direction, we have introduced $F_{S\parallel}({\bf q})$ for the $S^z$ component of spins and $F_{S\perp}({\bf q})$ for the $S^xS^y$-plane component. Also, since the breathing bond-alternation has already been incorporated in the spin Hamiltonian, we have taken ${\bf r}^0_i$ in Eqs. (\ref{eq:F_S}) as a regular position of the {\it uniform} pyrochlore lattice ignoring the bond-length alternation for simplicity. 

\begin{figure*}[t]
\begin{center}
\includegraphics[scale=0.85]{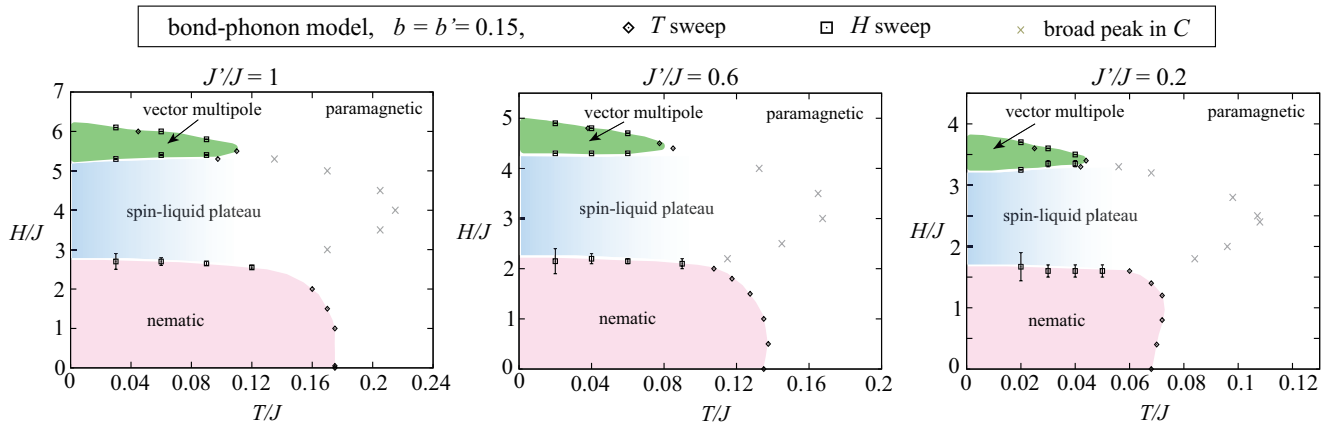}
\caption{Temperature and magnetic-field phase diagrams obtained in the bond-phonon model with $b=b'=0.15$ for $J'/J=1$ (left), $J'/J=0.6$ (center), and $J'/J=0.2$ (right). Note that a gray cross merely indicates a broad-peak temperature of the specific heat $C$ and thus, it does not mean a phase transition (for details, see the text). \label{fig:HT_bond}}
\end{center}
\end{figure*}

In both the bond-phonon and site-phonon models, once a spin state is obtained, the local lattice distortions can be evaluated directly from the given spin configuration \cite{Site_AK_16,Site_AK_19, HgCrO_Kimura_jpsj_14,CdCrO_Rossi_prl_19}, so that the essential information of the spin-lattice-coupled orders consists in the spin state. Thus, in this paper, we will focus basically on the ordering properties only of spins. By measuring the above physical quantities defined in Eqs. (\ref{eq:OP_nematic_div}) and (\ref{eq:F_S}) as well as the fundamental ones such as the specific heat $C = \frac{1}{T^2N}\big( \langle {\cal H}^2\rangle - \langle {\cal H} \rangle^2\big)$ and the magnetization $m = \langle | \frac{1}{N}\sum_{i} {\bf S}_i | \rangle$, we identify low-temperature phases in the applied magnetic field $H$.

\subsection{Numerical method}
To calculate the physical quantities introduced above, we perform Monte Carlo (MC) simulations for the bond-phonon and site-phonon Hamiltonians (\ref{eq:Hamiltonian_BP}) and (\ref{eq:Hamiltonian_SP}). Since our cubic unit cell contains 16 sites (see Fig. \ref{fig:snap_site015J06}), the total number of spins $N$ is related to the linear system size $L$ via $N=16 L^3$. In our MC simulation, we basically perform $2\times 10^6$ MC sweeps at each temperature and magnetic field with the periodic boundary condition, and the first half is discarded for thermalization. A single spin flip at each site consists of the conventional Metropolis update and a successive over-relaxation-like process in which we try to rotate a spin by the angle $\pi$ around the local mean field \cite{Loop_Shinaoka_14}. Observations are done in every 10 MC steps and the statistical average is taken over 4-8 independent runs. 
In most cases of the present system, efficient MC algorithms such as the temperature exchange method \cite{Fukushima_exchange} and a cluster update method, so-called loop-flip algorithm \cite{Loop_Shinaoka_14}, do not work except for some parameters (see below). Thus, we perform the single-spin-flip MC simulations, as mentioned above.

In the present single-spin-flip MC simulation, we often encounter various metastable states, as the spin state is not efficiently updated, being trapped in a local minimum. In particular, in the site-phonon model possessing inter-tetrahedron complex interactions, the low-temperature spin states obtained in the four different processes, cooling and warming runs at a fixed field and field-increase and field-decrease runs at a fixed temperature, sometimes differ. In such a situation, we compare the thermal-averaged values of the energy of these states and regard the lowest-energy state as the equilibrium state. Since this procedure could be applicable only to the lower-temperature region where the entropy effect is relatively weak, in relatively higher-temperature regions, we use the mixed-phase method (see Ref. \cite{MixedMethod_Creutz_79} and Appendix A) taking account of the entropy effect. Based on the above analysis, we determine the temperature and magnetic-field phase diagrams shown in Figs. \ref{fig:HT_bond} and \ref{fig:HT_siteall}, where the phase boundary between the low-temperature ordered and high-temperature disordered states is determined by the cooling runs. We note in passing that near the strong first-order transition between the low-field and middle-field $\frac{1}{2}$-plateau phases, we first use the mixed-phase method to obtain the equilibrium state, and then calculate the thermal average of the physical quantities (for details, see Appendix A). 

In addition to the local-update MC simulations, we use the temperature-exchange method \cite{Fukushima_exchange} to verify the spin-liquid behavior appearing in the $\frac{1}{2}$-plateau region in the bond-phonon model [see Fig. \ref{fig:bond} (c)] and the complex magnetic structures of the SP3 and SP4 phases in the site-phonon model shown in Fig. \ref{fig:snap_site020J02H0350-0370}. In the latter case, the temperature-exchange method can be applied to the $L=3$ system which is the smallest size for the SP3 and SP4 phases, but for the larger size of $L=6$, it does not work because of the first-order character of the transition from the high-temperature paramagnetic phase. 

Furthermore, we examine the ground state of the site-phonon model for the small number of spins $N=32$, and check that the results obtained in the finite-temperature MC simulations are consistent with those obtained in the ground-state analysis. To search for the global minimum of the Hamiltonian (\ref{eq:Hamiltonian_SP}), we use the ``NMinimize'' function in the Wolfram {\it Mathematica} software 11.3.0. 
With these multiple checks, we believe that the finite-temperature phases determined in the above procedure are the true equilibrium states, although we cannot rule out the possibility that there exists another phase which cannot be reached by any of the numerical methods used here.

Throughout this paper, we restrict ourselves to the case of $b=b'$ for simplicity, although in general, the SLC parameters $b$ and $b'$ defined in Eq. (\ref{eq:b-def}) should take different values in the breathing case.

\section{Result in the bond-phonon model}
In this section, we will discuss the ordering properties of the bond-phonon model.
In the bond-phonon model on the {\it uniform} pyrochlore lattice, it was shown by Shannon, Penc, and Motome that any magnetic (dipolar) LRO does not appear at any finite temperature and magnetic field, but instead, the two different quadruple LRO's, the nematic and vector-multipole orders each characterized by the nonzero value of $P_{\perp 2}$ and $P_{\perp 1}$, are realized at low and high fields, respectively, whereas at middle fields, the paramagnetic phase persists down to $T=0$ showing the spin-liquid behavior \cite{Bond_Shannon_10}. This spin-liquid state consists only of the 3-up and 1-down tetrahedra, so that it possesses the magnetization of $m=\frac{1}{2}$. Furthermore, the spin-liquid state with $m=\frac{1}{2}$ extends over the middle-field window, exhibiting the $\frac{1}{2}$ plateau in the magnetization curve, so that it is called the spin-liquid plateau phase. In this section, we will discuss the stabilities of the nematic, vector-multipole, and spin-liquid-plateau phases against the breathing bond-alternation, i.e., the change in $J'/J$.

\begin{figure}[t]
\begin{center}
\includegraphics[width=\columnwidth]{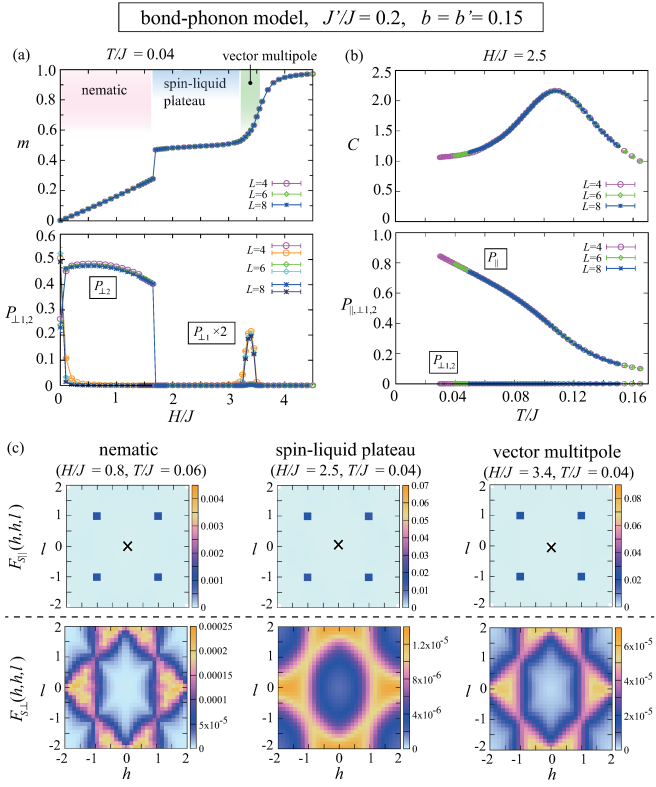}
\caption{MC results for the bond-phonon model with $b=b'=0.15$ and $J'/J=0.2$. (a) Field dependence of the magnetization $m$ (upper panel) and the $S^xS^y$-plane spin collinearities $P_{\perp 1}$ and $P_{\perp 2}$ (lower panel) at $T/J=0.04$. (b) Temperature dependence of the specific heat $C$ (upper panel) and the spin collinearities $P_{\parallel}$, $P_{\perp 1}$, and $P_{\perp 2}$ (lower panel) at $H/J=2.5$. (c) Spin structure factors $F_{S \parallel}({\bf q})$ (upper panels) and $F_{S\perp}({\bf q})$ (lower panels) in the $(h,h,l)$ plane obtained at $H/J=0.8$ and $T/J=0.06$ (left), $H/J=2.5$ and $T/J=0.04$ (center), and $H/J=3.4$ and $T/J=0.04$ (right) for $L=8$. In $F_{S \parallel}({\bf q})$, the high-intensity trivial peak at ${\bf q}=0$ indicated by the cross, which corresponds to $m^2$, has been removed.  \label{fig:bond}}
\end{center}
\end{figure}
Figure \ref{fig:HT_bond} shows the $J'/J$ dependence of the temperature and magnetic-field phase diagram in the bond-phonon model with $b=b'=0.15$. Although the characteristic temperature scales, e.g., the transition temperature between the paramagnetic and ordered phases, are suppressed with decreasing $J'/J$ (increasing the strength of the breathing bond-alternation), the relative stability among the nematic, vector-multipole, and spin-liquid-plateau phases is almost unchanged. Below, we will discuss the nature of these three phases.
 
In Fig. \ref{fig:bond}, we show the field dependence of various physical quantities in the strongly breathing case of $J'/J=0.2$. 
One can see from the upper panel of Fig. \ref{fig:bond} (a) that the magnetization $m$ increases linearly with increasing the applied field $H$ and, via a discontinuous first-order transition, it shows the $\frac{1}{2}$ plateau which is followed by the continuous growth in the higher-field phase. The low-field, middle-field $\frac{1}{2}$-plateau, and high-field phases correspond to the nematic, spin-liquid-plateau, and vector-multipole phases, respectively.     
As one can see from the $S^xS^y$-component spin structure factors $F_{S\perp}({\bf q})$ shown in the lower panels of Fig. \ref{fig:bond} (c), any magnetic Bragg reflections cannot be found in all the three phases, suggesting that the spin components perpendicular to the applied field are disordered. Such a situation is also the case for the $S^z$ spin component parallel to the field. Actually, in the $S^z$-component spin structure factors $F_{S\parallel}({\bf q})$ shown in the upper panels of Fig. \ref{fig:bond} (c), one cannot find nontrivial Bragg peaks except the $(1,1,1)$-type peaks originating from the uniform magnetization $m$. Thus, in all the three phases, spins (dipole moments) are still disordered down to the lowest temperature.

The low-temperature ordered and high-temperature paramagnetic phases can be distinguished by the spin collinearity. One can see from the lower panel of Fig. \ref{fig:bond} (a) that the low-field nematic and high-field vector-multipole phases are characterized by the nonzero values of $P_{\perp 2}$ and $P_{\perp 1}$, respectively. In middle-field $\frac{1}{2}$-plateau phase, on the other hand, a LRO does not occur for the quadruple moments as well as the dipole moments (spins), i.e., $P_{\perp 2}=P_{\perp 1}=0$, so that this phase is a spin-liquid state as is also suggested from the specific-heat data shown in Fig. \ref{fig:bond} (b) where a signature of a phase transition cannot be seen. Note that the broad peak in $C$ is associated with the growth of the collinearity along the field direction $P_{\parallel}$ [see the lower panel of Fig. \ref{fig:bond} (b)] which could also be interpreted as the formation of the 3-up and 1-down spin configuration on each tetrahedron. The broad-peak temperature is indicated by the gray cross in the temperature and magnetic-field phase diagrams in Fig. \ref{fig:HT_bond}.  

\section{Result in the site-phonon model}
In this section, we will discuss the ordering properties of the site-phonon model in which the inter-tetahedron interactions work as effective further neighbor interactions, leading to magnetic LRO's.

Since as mentioned in Sec. I, the zero-field phase realized in the weak SLC regime of the site-phonon model has the same spin structure as that observed in Li(Ga, In)Cr$_4$O$_8$ \cite{BrPyro_Nilsen_15,BrPyro_Saha_16,Site_AK_19}, we will focus on this weak SLC regime of $b=b'<0.25$ which should be relevant to the existing materials. This zero-field spin state, which is realized on both the uniform and breathing pyrochlore lattices, is tetragonal-symmetric being characterized by the $(1,1,0)$-type magnetic Bragg reflections.
Concerning the in-field properties of the site-phonon model on the {\it uniform} pyrochlore lattice, it was shown by Bergman {\it et al.} that the $\frac{1}{2}$ plateau shows up in the magnetization curve as in the case of the bond-phonon model and that the ground-state spin structure of the $\frac{1}{2}$-plateau phase is the $R$ state with the $P$4$_3$32 symmetry \cite{Site_Bergman_06} (the P2 phase explained below is exactly the same as the $R$ state) which consists of the $\uparrow\uparrow\uparrow\downarrow$ chains running along all the bond directions, keeping the 3-up and 1-down configuration on each tetrahedron. This $\uparrow\uparrow\uparrow\downarrow$ state is cubic symmetric, which is consistent with the experimental results on HgCr$_2$O$_4$ and CdCr$_2$O$_4$ \cite{HgCrO_Matsuda_07, CdCrO_Matsuda_prl_10}. Bearing these fundamental physics in our mind, we will discuss in-field properties of the site-phonon model on the {\it breathing} pyrochlore lattice. 

\subsection{Ground-state analysis}
\begin{figure*}[t]
\centering
\includegraphics[width=\linewidth]{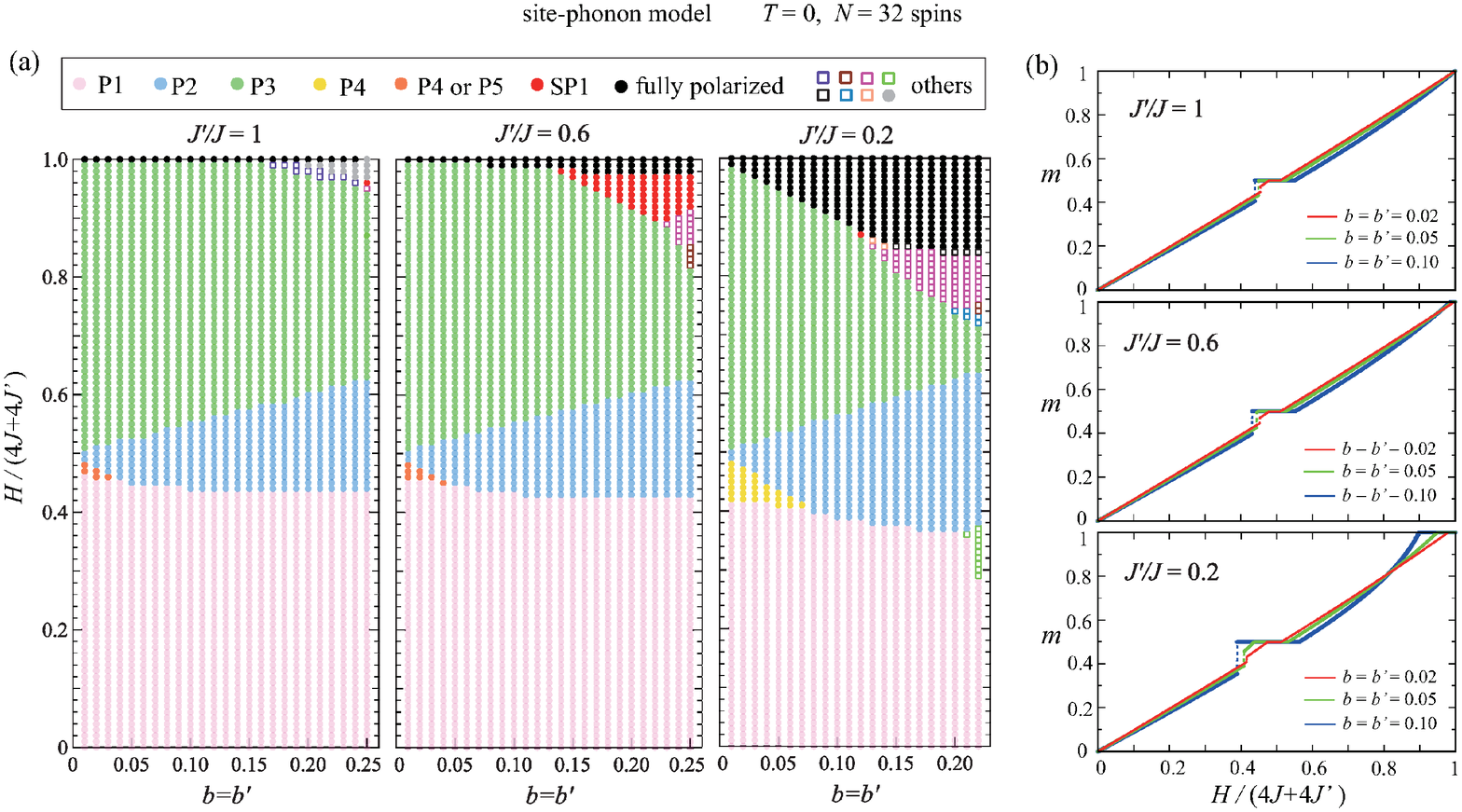}
\caption{Result of the ground-state analysis of the site-phonon model for the small number of $N=32$ spins. (a) SLC parameter $b=b'$ and the magnetic field $H$ dependence of the ground state for $J'/J=1$ (left), $J'/J=0.6$ (center), and $J'/J=0.2$ (right). Filled (open) symbols denote 16-sublattice (32-sublattice) magnetic structures. (b) Associated magnetization curves for $b=b=0.02$ (red), $b=b=0.05$ (green), and $b=b=0.10$ (blue) in the cases of $J'/J=1$ (top), $J'/J=0.6$ (middle), and $J'/J=0.2$ (bottom).}
\label{fig:GS}
\end{figure*}

We first discuss results of the ground-state analysis for $N=32$ spins. Since the zero-field ground state of the site-phonon model is a 16-sublattice state characterized by $(1,1,0)$ Bragg reflections or a 32-sublattice one characterized by $(\frac{1}{2},\frac{1}{2},\frac{1}{2})$ reflections \cite{Site_AK_19}, even the small number of spins $N=32$ can describe at least the zero-field magnetic structures. 
Figure \ref{fig:GS} (a) shows the SLC parameter $b=b'$ and the magnetic field $H$ dependence of the ground state for $J'/J=1$ (left), $J'/J=0.6$ (center), and $J'/J=0.2$ (right), where the magnetic field is normalized by the saturation field for $b=b'=0$, i.e., $4J+4J'$ \cite{BrPyro_Hdep_Gen_20}. Note that in the limit of $b=b' \rightarrow 0$, the ground state remains disordered even in the presence of a magnetic field at least in the uniform case of $J'/J=1$ \cite{Bond_Shannon_10}. For the parameter sets we investigated, we obtain 14 kinds of magnetic structures except for the trivial high-field fully-polarized state. A half of them are 16-sublattice states and the remaining half are 32-sublattice ones the stability regions of which are, respectively, represented by filled and open symbols in Fig. \ref{fig:GS} (a).

In the uniform case of $J'/J=1$, three magnetically long-range-ordered states are realized in a wide range of the parameters space, i.e., the low-field, middle-field, and high-field phases which, hereafter, will be called P1, P2, and P3 phases, respectively. These P1, P2, and P3 phases appear also in the breathing cases of $J'/J=0.6$ and $J'/J=0.2$. For smaller values of $b=b'$, an intermediate phase appears between the P1 and P2 phases in both the uniform and breathing cases [see the orange and yellow regions in Fig. \ref{fig:GS} (a)]. This intermediate phase is the cant 2:1:1 state \cite{Bond_Penc_04,Bond_Shannon_10} or the ``1-up, 1-down, and V'' state which will be called P4 and P5 phases, respectively (for their real-space structures, see Fig. \ref{fig:cant211U} in Appendix B). The P4 phase is favored in the strongly breathing case of $J'/J=0.2$, whereas in the uniform and weakly breathing cases of $J'/J=1$ and 0.6, the P4 phase is degenerate with the P5 phase at least within our computation accuracy. 

For moderate SLC, on the other hand, additional new phases are favored by the breathing bond-alternation. For example, in the weakly breathing case of $J'/J=0.6$, there exists a higher-field phase between the saturation field and the P3 phase which we call SP1 phase. The SP1 phase is the 16-sublattice state, similarly to the P1, P2, P3, P4, and P5 phases. In the relatively large $b$ region, a variety of 32-sublattice states become favorable near the saturation field and the $\frac{1}{2}$ plateau, reflecting the fact that in the strong SLC regime of $b,b'>0.25$, the zero-field ground state is the $(\frac{1}{2},\frac{1}{2},\frac{1}{2})$-type 32-sublattice state.

Figure \ref{fig:GS} (b) shows the parameter dependence of the magnetization curve. One can see that with increasing the SLC parameter $b=b'$, the $\frac{1}{2}$ plateau, which corresponds to the P2 phase, gets wider and the associated magnetization jump from the P1 phase becomes more remarkable. The existence of the P4 or P5 phase is reflected as the bending of the plateau just above the magnetization jump. We note that although the top and middle panels of Fig. \ref{fig:GS} (b) are obtained by assuming that the intermediate phase is the P5 phase, the magnetization curve is not altered even if it is assumed to be the P4 phase. 

Although the above ground-state analysis for $N=32$ spins offers crucial information about magnetic LRO's in the site-phonon model, careful analysis is necessary to check whether the states obtained for $N=32$ are really stable or not in the thermodynamic limit of $L \rightarrow \infty$, or equivalently, $N \rightarrow \infty$. In particular, in the relatively large $b$ and smaller $J'/J$ regions, the occurrence of the 32-sublattice states in the ground-state analysis indicates that a large number of possibly more than 32 spins should be taken into account to lower the energy of the system.
Indeed, as we will see below, in the finite-temperature MC simulations, the tetrahedron-based orders involving more than 32 spins (SP2, SP2', SP3, and SP4 phases) appear just below the saturation field and the $\frac{1}{2}$ plateau for $J'/J=0.2$. In addition, the P3 phase is slightly modified in the strongly breathing case $J'/J=0.2$.

\subsection{Temperature and magnetic-field phase diagram}
\begin{figure*}[t]
\begin{center}
\includegraphics[scale=0.85]{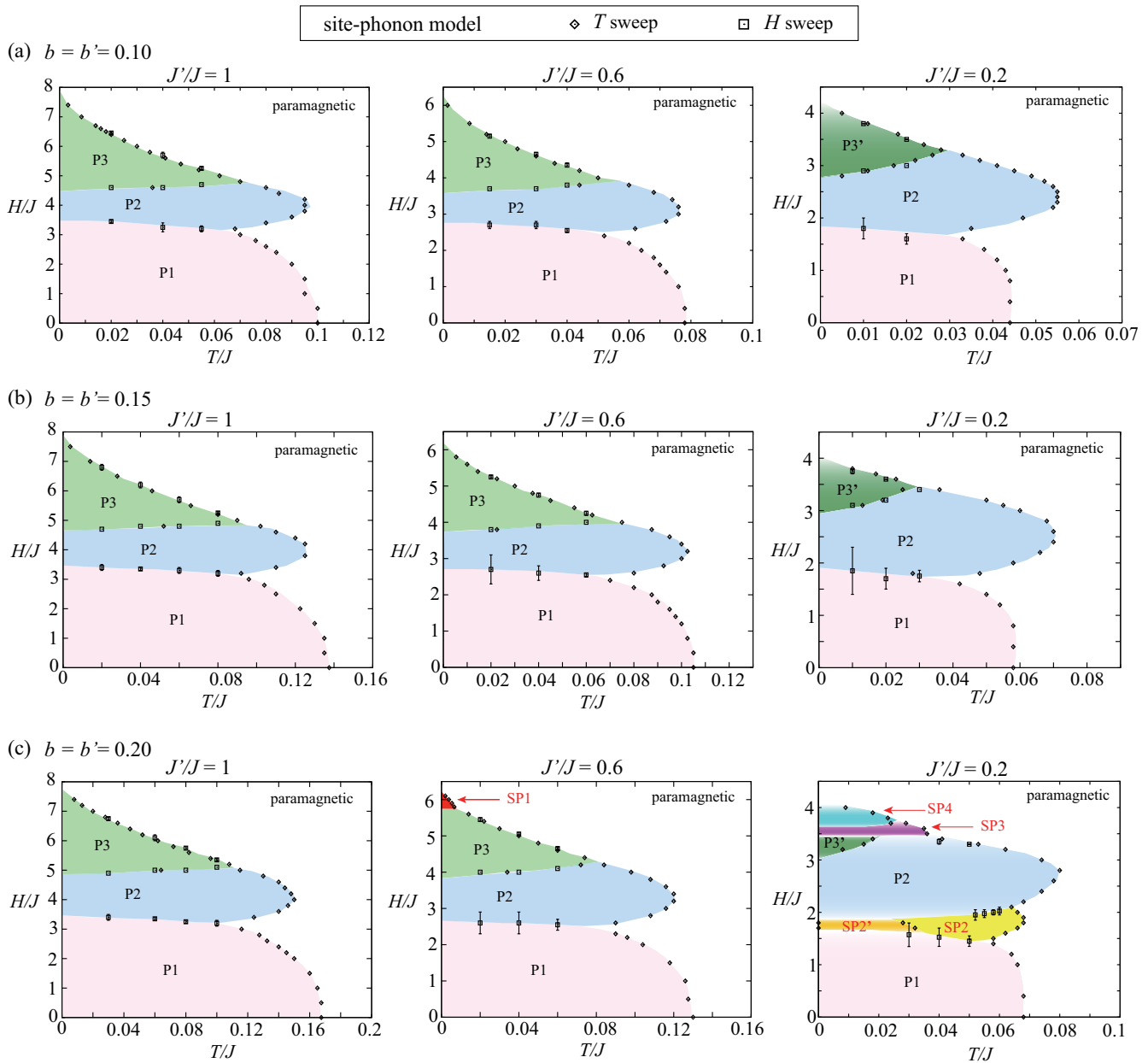}
\caption{Temperature and magnetic-field phase diagrams obtained in the site-phonon model with (a) $b=b'=0.10$, (b) $b=b'=0.15$, and (c) $b=b'=0.20$ for $J'/J=1$ (left), $J'/J=0.6$ (center), and $J'/J=0.2$ (right). \label{fig:HT_siteall}}
\end{center}
\end{figure*}

Now that we have understood the ground-state properties of the site-phonon model for $N=32$, we will next discuss finite-temperature properties for larger numbers of spins $N=16L^3$ with $L \geq 4$. Since the characteristic in-field feature of the spin-lattice-coupled system is the occurrence of the $\frac{1}{2}$ plateau, hereafter, we will focus on the SLC parameters of $b=b' \geq 0.10$ for which the $\frac{1}{2}$ plateau is relatively wide.  

Figure \ref{fig:HT_siteall} shows the $J'/J$ dependence of the temperature and magnetic-field phase diagram in the site-phonon model with the SLC parameters of $b=b'=0.10$, 0.15, and 0.20. The finite-temperature results in Fig. \ref{fig:HT_siteall} are basically consistent with the results obtained in the ground-state analysis for the small number of spins $N=32$ [see Fig. \ref{fig:GS} (a)] except that in the strongly breathing case of $J'/J=0.2$, LRO's involving more than 32 spins appear as the lowest energy state. We note that according to the ground-state analysis for $b=b'=0.15$ and $J'/J <1$, an additional higher-field phase may exist just above the P3 or P3' phase in the center and right panels of Fig. \ref{fig:HT_siteall} (b), but it is not obtained in the temperature range of our MC simulations. Before going to the details of the ordered phases, here, we will briefly summarize the result.

In the uniform case of $J'/J=1$, the three magnetically long-range-ordered states, P1, P2, and P3 phases are stabilized in the low-field, middle-field, and high-field regions, respectively. The P1 and P2 phases are robust against the breathing alternation, while the P3 one is not. Although the P3 phase still exists in the weakly breathing case of $J'/J=0.6$, its spin structure is modified in the strongly breathing case of $J'/J=0.2$. We call such a modified state P3' phase. For the weak SLC's of $b=b'=0.10$ and 0.15, only the above four phases, P1, P2, P3, and P3', come into play. By contrast, for the moderate SLC of $b=b'=0.20$, additional new phases are induced by the breathing bond-alternation. In the weakly breathing case of $J'/J=0.6$, the SP1 phase appears in a higher-field region, whereas in the strongly breathing case of $J'/J=0.2$, there are two corresponding higher-field phases which will be called SP3 and SP4 phases. Another noteworthy aspect in the case of $J'/J=0.2$ is the occurrence of the intermediate phase between the P1 and P2 phases. This phase just below the $\frac{1}{2}$ plateau will be named SP2 phase and its lower-temperature state is SP2' phase in which the correlation between the $S^xS^y$ components of spins is different from that in the higher-temperature SP2 phase. 
Now, we shall discuss the details of the ordered states, starting from the P1, P2, and P3 phases appearing on both the {\it uniform} and {\it breathing} pyrochlore lattices.

\subsection{Ordered states appearing on both the uniform and breathing pyrochlore lattices} 
\begin{figure}[t]
\begin{center}
\includegraphics[scale=0.74]{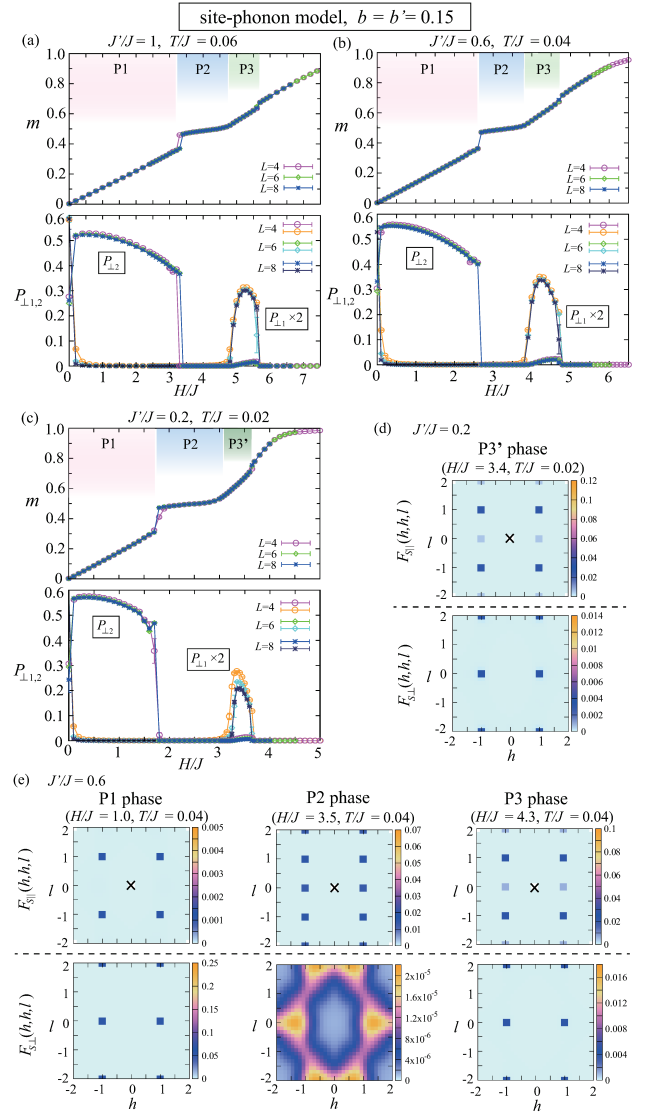}
\caption{MC results for the site-phonon model with $b=b'=0.15$. (a)-(c) Field dependence of $m$, $P_{\perp 1}$, and $P_{\perp 2}$ for (a) $J'/J=1$ and $T/J=0.06$, (b) $J'/J=0.6$ and $T/J=0.04$, and (c) $J'/J=0.2$ and $T/J=0.02$. (d) Spin structure factor obtained in the high-field P3' phase at $H/J=3.4$ and $T/J=0.02$ for $L=8$ in the strongly breathing case of $J'/J=0.2$. (e) Spin structure factors  obtained at $T/J=0.04$ in the weakly breathing case of $J'/J=0.6$. The left, center, and right panels are respectively obtained in the low-field P1 phase at $H/J=1.0$, the middle-field P2 phase at $H/J=3.5$, and the high-field P3 phase at $H/J=4.3$ for $L=8$. In $F_{S \parallel}({\bf q})$, the high-intensity trivial peak at ${\bf q}=0$ indicated by the cross, which corresponds to $m^2$, has been removed.
\label{fig:site015}}
\end{center}
\end{figure}
Figure \ref{fig:site015} shows the field dependence of various physical quantities in the uniform ($J'/J=1$), weakly breathing ($J'/J=0.6$), and strongly breathing ($J'/J=0.2$) cases for $b=b'=0.15$. As readily seen from the upper panels of Figs. \ref{fig:site015} (a), (b), and (c), the $\frac{1}{2}$ plateau shows up in the magnetization curve being relatively robust against the breathing bond-alternation, i.e., the change in $J'/J$. This $\frac{1}{2}$-plateau phase is the P2 phase, and the low-field (high-field) phase below (above) the $\frac{1}{2}$ plateau is the P1 (P3 or P3') phase. As one can see from the lower panels of Figs. \ref{fig:site015} (a) and (b), the P1 and P3 phases are characterized by nonzero values of $P_{\perp 2}$ and $P_{\perp 1}$, respectively. These features of the low-field (P1), middle-field (P2), and high-field (P3) phases are quite similar to those of the nematic, spin-liquid-plateau, and vector-multipole phases in the bond-phonon model [see Fig. \ref{fig:bond} (a)], but in the present site-phonon model, magnetic LRO's are realized in the P1, P2, and P3 phases, as is indicated by Bragg peaks in the spin structure factors $F_{S\perp}({\bf q})$ and $F_{S\parallel}({\bf q})$ shown in Fig. \ref{fig:site015} (e). 

In the low-field P1 phase, $F_{S\perp}({\bf q})$ for the $S^xS^y$ spin component perpendicular to the field exhibits Bragg peaks at $\pm(1,1,0)$, while $F_{S\parallel}({\bf q})$ for the $S^z$ spin component parallel to the field only exhibits the trivial $(0,0,0)$ and $(1,1,1)$ peaks stemming from the uniform magnetization [see the left panels of Fig. \ref{fig:site015} (e)]. The P1 phase characterized by the $(1,1,0)$-type Bragg reflections is tetragonal-symmetric in the sense that among the three equivalent points $(1,1,0)$, $(1,0,1)$, and $(0,1,1)$, only one is selected. The real-space spin configuration of the P1 phase is shown in Fig. \ref{fig:snap_site015J06} (a). As one can see from the periodic pattern of the yellow and red arrows in Fig. \ref{fig:snap_site015J06} (a), the perpendicular spin components constitute $\uparrow\downarrow\uparrow\downarrow$ chains along the facing two bonds of a tetrahedron and $\uparrow\uparrow\downarrow\downarrow$ chains along the rest four bonds. All the spins are canted along the field direction. In units of tetrahedron, each of all the tetrahedra takes the cant 2:2 spin configuration. 

In the middle-field P2 phase characterized by the $\frac{1}{2}$-magnetization plateau, as one can see from the real-space spin configuration shown in Fig. \ref{fig:snap_site015J06} (b), the parallel spin components constitute $\uparrow\uparrow\uparrow\downarrow$ chains along all the six tetrahedral bonds, keeping the 3-up and 1-down configuration on each tetrahedron. This $\uparrow\uparrow\uparrow\downarrow$ chain structure is reflected in $F_{S\parallel}({\bf q})$ as the magnetic Bragg peaks at $(1,1,0)$ [see the center upper panel of Fig. \ref{fig:site015} (e)] and other cubic-symmetric points of $(0,1,1)$ and $(1,0,1)$. Thus, the spin state is cubic-symmetric. In the P2 phase, all the spins are collinearly aligned along the field direction, and the $S^xS^y$ components perpendicular to the field only exhibit a short-range correlation. Actually, any Bragg reflections cannot be found in the $S^xS^y$-component spin structure factor $F_{S\perp}({\bf q})$ shown in the center lower panel of Fig. \ref{fig:site015} (e). 

The high-field P3 phase is a canted state of the P2 phase. As one can see from the real-space configuration shown in Fig. \ref{fig:snap_site015J06} (c), the $\uparrow\uparrow\uparrow\downarrow$ chains are canted from the field direction such that the cant directions of the up and down spins are antiparallel to each other keeping the $S^xS^y$ components to be collinear. As a result, not only the $S^z$ component but also the $S^xS^y$ one constitutes $\uparrow\uparrow\uparrow\downarrow$ chains running along all the six tetrahedral bonds. Reflecting this ordering pattern, the spin structure factor shown in the right panels of Fig. \ref{fig:site015} (e) has the $(1,1,0)$-type Bragg peaks not only in the $S^z$ sector [$F_{S\parallel}({\bf q})$] but also in the $S^xS^y$ sector [$F_{S\perp}({\bf q})$]. In contrast to the P2 phase, the $S^xS^y$ components exhibit the LRO in the P3 phase. Comparing the lower left and right panels of Fig. \ref{fig:site015} (e), one notices that $F_{S\perp}({\bf q})$'s of the P1 and P3 phases look quite similar to each other, but in the P3 phase, the Bragg peaks show up at the wave vectors of $(1,0,1)$ and $(0,1,1)$ as well as $(1,1,0)$. Thus, the P3 phase is cubic-symmetric. 
Although the P3 phase is robust against the weak breathing bond-alternation of $J'/J=0.6$, it is modified into the P3' phase in the strongly breathing case of $J'/J=0.2$.

\begin{figure}[t]
\begin{center}
\includegraphics[scale=0.55]{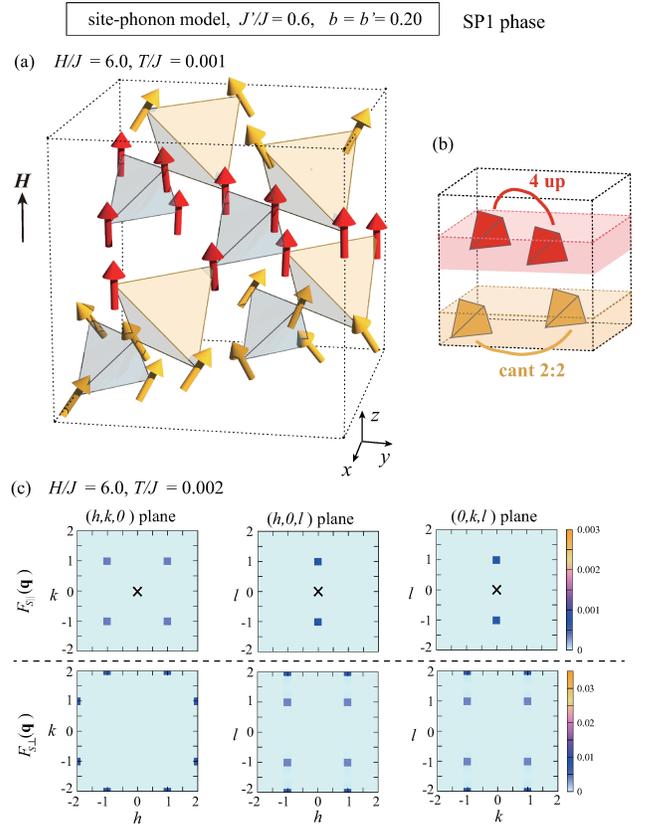}
\caption{MC results obtained in the higher-field SP1 phase at $H/J=6.0$ for the site-phonon model with $b=b'=0.20$ and $J'/J=0.6$. (a) Spin snapshot at $T/J=0.001$ and (b) the associated small-tetrahedron distribution within the cubic unit cell. (c) Spin structure factors $F_{S \parallel}({\bf q})$ (upper panels) and $F_{S\perp}({\bf q})$ (lower ones) in the $(h,k,0)$ (left), $(h,0,l)$ (center), and $(0,k,l)$ (right) planes obtained at $T/J=0.002$ for $L=8$. In $F_{S \parallel}({\bf q})$, the high-intensity trivial peak at ${\bf q}=0$ indicated by the cross has been removed.   
\label{fig:snap_site020J06}}
\end{center}
\end{figure}

\subsection{Ordered states favored on the breathing pyrochlore lattice} 
Among various magnetic LRO's induced by the breathing lattice structure, we first discuss the P3' phase. As shown in Figs. \ref{fig:site015} (c) and (d), the ordering properties of the high-field P3' phase are similar to those of the P3 phase: the $(1,1,0)$-type magnetic Bragg peaks can be found in both $F_{S\parallel}({\bf q})$ and $F_{S\perp}({\bf q})$, and the in-plane collinearity $P_{\perp 1}$ develops on entering the P3' phase from the P2 phase. When we take a closer look at the system-size dependence of $P_{\perp 1}$, however, $P_{\perp 1}$ in the P3' phase is suppressed with increasing the system size $L$, which is in sharp contrast to the corresponding behavior in the P3 phase [see the lower panels of Figs. \ref{fig:site015} (a) and (b)]. We note that in the MC simulation, even when we take the spin configuration of the P3 phase as the initial state, the system eventually goes to a slightly disturbed state, and $P_{\perp 1}$ is suppressed to be a smaller value than those in the P1 and P3 phases. Furthermore, the energy of the system gradually becomes lower than that of the P3 phase with increasing the system size $L$, although the Bragg peak is still located at $(1,1,0)$ up to the largest size of $L=12$. Thus, there is a possibility that in the P3' phase, an incommensurate order whose wave vector is close to $(1,1,0)$ might be realized in the thermodynamic limit of $L\rightarrow\infty$. Identifying the spin structure of the P3' phase needs larger-$L$ analysis, but we will leave this issue for our future work. 

The above P1, P2, P3, and P3' phases are realized in all the three cases of $b=b'=0.10$, 0.15, and 0.20 (see Fig. \ref{fig:HT_siteall}). For the relatively strong SLC of $b=b'=0.20$, additional new phases, SP1, SP2, SP2', SP3, and SP4 phases, are induced by the breathing bond-alternation. Below, we will discuss these phases starting from the SP1 phase occurring just below the saturation field in the weakly breathing case of $J'/J=0.6$ [see the center panel of Fig. \ref{fig:HT_siteall} (c)].

Figure \ref{fig:snap_site020J06} shows the spin structure of the SP1 phase. One can see from the real-space spin configuration shown in Fig. \ref{fig:snap_site020J06} (a) that the SP1 phase is the 16-sublattice state similarly to the P1, P2, and P3 phases. In contrast to the P1, P2, and P3 phases where all the tetrahedra are equivalent to one another taking the same spin configuration on each tetrahedron, the SP1 phase consists of the two different types of small tetrahedra, 4-up and cant 2:2 tetrahedra. As one can see from Figs. \ref{fig:snap_site020J06} (a) and (b), the distribution of the 4-up and cant 2:2 tetrahedra within the cubic unit cell is tetragonal-symmetric: the 4-up tetrahedron pair and the cant 2:2 tetrahedron pair are stacking along the $z$ axis, leading to the alternating stack of the 4-up and cant 2:2 layers in the whole system. The tetragonal symmetry of this kind can clearly be seen in the difference in the spin structure factors in the $(h,k,0)$, $(h,0,l)$, and $(0,k,l)$ planes. As readily seen from Fig. \ref{fig:snap_site020J06} (c), among the cubic-symmetric families of $(1,0,0)$, $(0,1,0)$, and $(0,0,1)$, only $(0,0,1)$ is picked up in $F_{S\perp}({\bf q})$ and such a situation is also the case for the family of $(0,1,1)$, $(1,0,1)$, and $(1,1,0)$. 
We note that although in the case of Fig. \ref{fig:snap_site020J06}, the $z$-direction is special, any of the $x$, $y$, and $z$ directions can be the tetragonal axis, i.e., the stacking direction of the alternating 4-up and cant 2:2 layers. Actually, the stacking direction differs run to run in the MC simulations.  

\begin{figure}[t]
\begin{center}
\includegraphics[scale=0.52]{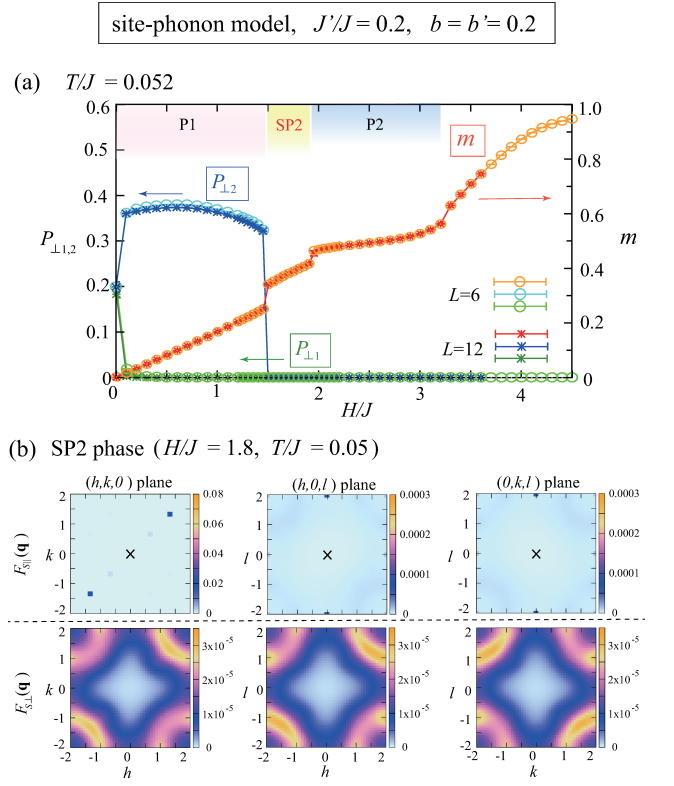}
\caption{ (a) The field dependence of $m$ (reddish symbols), $P_{\perp 1}$ (greenish ones), and $P_{\perp2}$ (bluish ones) at $T/J=0.052$ for the site-phonon model with $b=b'=0.20$ in the strongly breathing case of $J'/J=0.2$. The SP2 phase is realized in the field range of $1.5 \leq H/J \leq 1.9$. (b) Spin structure factor obtained at $H/J=1.8$ and $T/J=0.05$ for $L=12$, where the notations are the same as those of Fig. \ref{fig:snap_site020J06} (c). \label{fig:site020J02H0180}}
\end{center}
\end{figure}

In the strongly breathing case of $J'/J=0.2$, as shown in the right panel of Fig. \ref{fig:HT_siteall} (c), the SP2 and SP2' (SP3 and SP4) phases appear just below the $\frac{1}{2}$ plateau (the saturation field). 
Figure \ref{fig:site020J02H0180} (a) shows the field dependence of the magnetization $m$ and the $S^xS^y$-plane collinearities $P_{\perp 1}$ and $P_{\perp 2}$, where the SP2 phase is stabilized in the field range of $1.5 \leq H/J \leq 1.9$. As one can see from a two-step jump in the magnetization curve, the SP2 phase is separated by the first-order transitions from the low-field P1 and middle-field P2 phases. Figure \ref{fig:site020J02H0180} (b) shows the spin structure factor in the SP2 phase. $F_{S\parallel}({\bf q})$ exhibits the Bragg peaks at $\pm(\frac{2}{3},\frac{2}{3},0)$ and $\pm(\frac{4}{3},\frac{4}{3},0)$ but not other cubic-symmetric points, so that the state should be tetragonal. The $S^xS^y$ components of spins, on the other hand, are not long range ordered, as is suggested from the absence of Bragg peaks in $F_{S\perp}({\bf q})$ [see the lower panels of Fig. \ref{fig:site020J02H0180} (b)]. Together with the result that both $P_{\perp 1}$ and $P_{\perp 2}$ vanish in the SP2 phase [see Fig. \ref{fig:site020J02H0180} (a)], it turns out that the $S^xS^y$ components of spins remain disordered. We note that in the lower-temperature SP2' phase, the perpendicular components of spins are ordered into a state characterized by the nonzero value of $P_{\perp 2}$, and associated $(\frac{2}{3},\frac{2}{3},0)$-type Bragg peaks develop in $F_{S\perp}({\bf q})$ with $F_{S\parallel}({\bf q})$ being almost unchanged from that in the SP2 phase. 

The real-space structure of the SP2 phase is very complicated; it consists of as many as $16\times6^3$ spins. In units of tetrahedron, similarly to the SP1 phase, the SP2 phase is composed of the two different kinds of small tetrahedra, the cant 2:2 tetrahedra and the 3-up and 1-down ones, reflecting the fact that this phase is intercalated between the P1 and P2 phases each consisting of the cant 2:2 tetrahedra and the 3-up and 1-down ones, respectively. In the SP2 phase, these two elements, i.e., the cant 2:2 tetrahedra and the 3-up and 1-down ones, are arranged periodically over the whole system, forming a tetragonal-symmetric pattern (for details, see Fig. \ref{fig:snap_site020J02H0180} in Appendix C). 

The SP3 and SP4 phases appearing just below the saturation field are also tetrahedron-based orders. In these phases, the fundamental element is the 4-up tetrahedron. The SP3 and SP4 phases are characterized by different ordering patterns of the 4-up tetrahedra both of which are noncubic (see Fig. \ref{fig:snap_site020J02H0350-0370} in Appendix C).

Although the real-space structures of the SP2, SP2', SP3, and SP4 phases are rather complicated, the existence of nonequivalent tetrahedra is the common feature of these phases and the SP1 phase all of which are induced by the breathing bond-alternation. By contrast, in the P1, P2, and P3 phases appearing in both the uniform and breathing pyrochlore lattices, all the tetrahedra are equivalent, having the same spin configuration. The key role of the breathing lattice structure is the installation of the nonequivalent tetrahedra in the system, leading to the tetrahedron-based magnetic orders.     

\section{Summary and Discussion}
In this paper, we have theoretically investigated in-field properties of the spin-lattice-coupled ordering in the breathing-pyrochlore antiferromagnets, based on the simplified models taking account of the effect of local lattice distortions, the so-called bond-phonon and site-phonon models. It is found by means of MC simulations that the site-phonon model exhibits a rich variety of magnetic LRO's some of which are unique to the breathing system, while the bond-phonon model does not show any magnetic (dipolar) LRO regardless of whether the system is breathing or not. In the site-phonon model with the small SLC parameters of $0.10 \leq b=b' \leq 0.20$, the low-field, middle-field, and high-field phases appearing in the {\it uniform} pyrochlore lattice, which are, respectively, named the P1, P2, and P3 phases in Fig. \ref{fig:HT_siteall}, are also realized on the breathing pyrochlore lattice. Note that the magnetization curve shows the $\frac{1}{2}$ plateau in the middle-field P2 phase. In addition to the three phases, as a result of the combined effect of the breathing bond-alternation which introduces the nonequivalent tetrahedra in the system and the inter-tetrahedron interactions characteristic of the site-phonon model, tetrahedron-based new LRO's are induced just below the $\frac{1}{2}$ plateau and the saturation field (SP1, SP2, SP2', SP3, and SP4 phases in Fig. \ref{fig:HT_siteall}). It is also found from the ground-state analysis that in the much weaker SLC regime where the $\frac{1}{2}$ plateau is quite narrow, an intermediate phase appears between the P1 and P2 phases. This intermediate phase is the P4 or P5 phase whose local spin configurations on each tetrahedron are the ``cant 2:1:1'' and ``1-up, 1-down and V'', respectively (see Fig. \ref{fig:GS} together with Fig. \ref{fig:cant211U} in Appendix B). 

In deriving the site-phonon effective spin Hamiltonian, the following approximations have been made: the exchange interaction is assumed to depend only on the distance between the two spins although the situation in real materials would be more complicated, and is expanded with respect to the site displacement up to the first order, neglecting the higher-order contributions [see Eqs. (\ref{eq:original_H}) and (\ref{eq:expansion})]. Also, the elastic energy of the lattice is assumed to be a local one depending only on each site, although in reality neighboring displacements should be correlated in the form of the dispersive phonon modes. Due to the assumed local nature of the phonons, physics relevant to the lattice such as the net lattice distortions, the volume changes, and the phonon dispersions cannot be described in the present model. Nevertheless, as will be discussed below, the site-phonon model captures the essential feature of the ''spin'' ordering physics of the spin-lattice-coupled phenomena in the chromium oxides not only at zero field but also at finite fields.
 
In experiments on the {\it uniform} pyrochlore antiferromagnets $A$Cr$_2$O$_4$ ($A$=Hg, Cd, Zn, Mg), the $\frac{1}{2}$ plateau has commonly been observed and its spin structure suggested from the neutron diffraction patterns for HgCr$_2$O$_4$ and CdCr$_2$O$_4$\cite{HgCrO_Matsuda_07, CdCrO_Matsuda_prl_10} is the same as that of the above P2 phase, i.e., the state consisting of $\uparrow\uparrow\uparrow\downarrow$ chains running along all the tetrahedral bonds. We note that although the P2 phase was already predicted by Bergman {\it et al} \cite{Site_Bergman_06}, its finite-temperature properties are clarified in the present paper. In addition to the P2 phase, the existence of the P4 or P5 phase just below the $\frac{1}{2}$ plateau is consistent with the observation of an intermediate phase in ZnCr$_2$O$_4$ \cite{ZnCrO_Miyata_jpsj_11,ZnCrO_Miyata_prl_11,ZnCrO_Miyata_jpsj_12} and MgCr$_2$O$_4$ \cite{MgCrO_Miyata_jpsj_14}. The observed intermediate phase has been interpreted as the cant 2:1:1 state based on the bond-phonon picture, but our result suggests that the P5 phase shown in Fig. \ref{fig:cant211U} (b) is also a possible candidate for this phase. At zero field, on the other hand, the P1 phase does not seem to be reported in $A$Cr$_2$O$_4$ \cite{CdCrO_Chung_05,ZnCrO_Lee_08,HgCrO_Matsuda_07,MgCrO_Ortega_08} where the spin-ordering patterns vary from material to material, indicating that effects beyond the present simplified model such as higher-order contributions in Eq. (\ref{eq:expansion}), dispersive phonon modes, and other interactions should also be relevant. Indeed, the additional DM interaction slightly modifies the P1 phase and the modified state is consistent with the Neel state of CdCr$_2$O$_4$ \cite{SLC_Chern_06}. These results suggest that the zero-field and in-field properties of the uniform pyrochlore antiferromagnets $A$Cr$_2$O$_4$ can be basically described by the site-phonon model in spite of the simplification of the lattice distortions. 

In the {\it breathing} pyrochlore antiferromagnets Li(Ga, In)Cr$_4$O$_8$, the P1 phase (the tetragonal-symmetric spin structure consisting of the 2-up and 2-down tetrahedra) is realized at zero field \cite{BrPyro_Nilsen_15,BrPyro_Saha_16,Site_AK_19}, so that the site-phonon model could also be applied to this class of magnets.
Accordingly, the $\frac{1}{2}$ plateau observed in LiInCr$_4$O$_8$ \cite{BrPyro_Hdep_Okamoto_17,BrPyro_Hdep_Gen_19} would point to the realization of the P2 phase (the cubic-symmetric spin structure consisting of the 3-up and 1-down tetrahedra).
Notably, the magnetization jump just below the $\frac{1}{2}$ plateau in LiInCr$_4$O$_8$ is drastic and amounts to $\sim$0.75~$\mu_{\rm B}$ \cite{BrPyro_Hdep_Gen_19}, which is comparable to that in HgCr$_2$O$_4$ with the strongest SLC among the {\it A}Cr$_2$O$_4$ family \cite{HgCrO_Kimura_jpsj_14, CdCrO_Kimura_jpsj_15, ZnCrO_Miyata_jpsj_11, MgCrO_Miyata_jpsj_14}.
This indicates that LiInCr$_4$O$_8$ possesses a relatively strong SLC. In our present theoretical work, we have demonstrated that in the site-phonon model with the relatively strong SLC of $b=b'=0.20$, the intermediate phase (the SP2 or SP2' phase) appears between the P1 and P2 phases in the strongly breathing case of $J'/J=0.2$, showing a two-step magnetization jump [see Fig. \ref{fig:site020J02H0180} (a)]. However, such a feature is not observed in LiInCr$_4$O$_8$, indicating that $b$ and/or $b'$ may be smaller than 0.20 in this compound. On the other hand, according to the recent first-principle calculations \cite{FirstPrinciple_Ghosh_npj_19}, the breathing parameter $J'/J$ in LiInCr$_4$O$_8$, which was at first estimated to be $\sim 0.1$ based on the empirical relationship between the NN Cr-Cr bond length and the exchange coupling \cite{BrPyro_Okamoto_13}, is highly temperature dependent and becomes close to 1 at 20~K slightly above the transition temperature \cite{FirstPrinciple_Ghosh_npj_19}. Thus, at present, it is not clear whether the absence of the intermediate phase in LiInCr$_4$O$_8$ is due to the SLC parameters $b$ and $b'$ or the breathing parameter $J'/J$.
Concerning the Ga compound, both the empirical and the first-principle calculations show that the ratio between $J$ and $J'$ is $\sim 0.6$, but the latter predicts that $J$ is smaller than $J'$ in contrast to the naive expectation \cite{FirstPrinciple_Ghosh_npj_19}. We believe that future high-field measurements on LiGaCr$_4$O$_8$ should provide fundamental information on the system parameters. 

In this work, we have assumed $b=b'$ for simplicity, but different values of the SLC parameters $b \neq b'$ may lead to various types of tetrahedron-based LRO's other than the SP1, SP2, SP2', SP3, and SP4 phases depending on the value of $J'/J$.
Although it might not be so easy to find where a complicated real material is placed in the parameter space of the site-phonon model, this work presenting the result for the simplified case of $b=b'$ should help the understanding of the spin-lattice-coupled orderings in the breathing pyrochlore antiferromagnets.

Finally, we would like to comment on the validity of the site-phonon model in other related systems. 
As pointed out in Ref.~\cite{BrPyro_Hdep_Gen_20}, the present model Eq.~(\ref{eq:Hamiltonian_SP}) on the breathing pyrochlore lattice can be extended to the case of antiferromagnetic $J>0$ and ferromagnetic $J'<0$. In this case, the cant 2:1:1 phase appears in a relatively wide parameter region, which seems to be relevant to the gradual magnetization increase prior to the $\frac{1}{2}$ plateau observed in CuInCr$_4$S$_8$ \cite{BrPyro_Sulfides_Okamoto_18, BrPyro_Hdep_Gen_20}.
Furthermore, the site-phonon model on the {\it triangular} lattice can reproduce the zero-field zigzag ground state and the in-field $\frac{1}{5}$-magnetization plateau observed in the multiferroic compound CuFeO$_2$ \cite{Site_Wang_08}. Hence, it would be intriguing to apply the site-phonon model to other frustrated spin systems to search for novel spin-lattice-coupled phenomena.

\begin{acknowledgments}
The authors thank R. Osamura for useful discussions. We are thankful to ISSP, the University of Tokyo and YITP, Kyoto University for providing us with CPU time. This work is supported by JSPS KAKENHI Grant Number JP16K17748, JP17H06137 and JP20J10988.
\end{acknowledgments}

\appendix

\section{Mixed-phase method}
\begin{figure}[t]
\begin{center}
\includegraphics[scale=0.8]{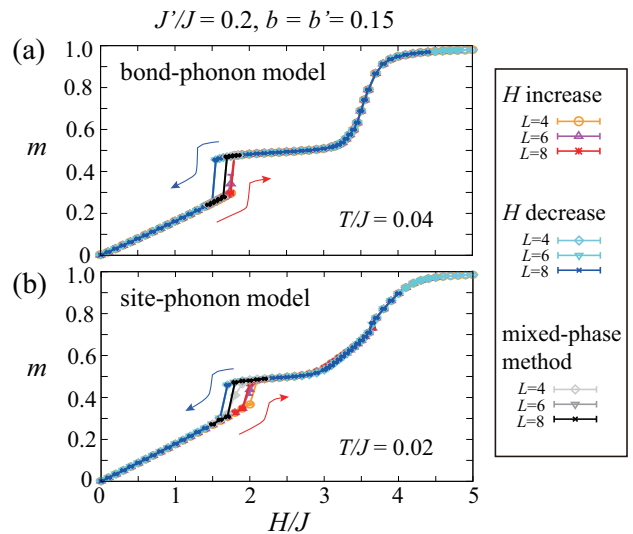}
\caption{Field dependence of the magnetization $m$ obtained in (a) the bond-phonon model and (b) the site-phonon model with $b=b'=0.15$ and $J'/J=0.2$. Reddish (Bluish) symbols denote the results obtained in the field-increase (field-decrease) process at a fixed temperature, and gray and black ones are obtained by the mixed-phase method. \label{fig:mixed}}
\end{center}
\end{figure}
When a first-order transition occurs between two phases, it is sometimes difficult to identify the phase boundary in the MC simulation because of a strong hysteresis. In such a situation, the mixed-phase method \cite{MixedMethod_Creutz_79} is useful. In the mixed-phase method, one first prepares an initial state in which the half of the system is occupied with a spin configuration of one phase and the rest with a spin configuration of the other competing phase, and then, carries out MC sweeps, tracing the state change. From the final state after the MC sweeps, one can determine which phase is more stable at a given temperature and a magnetic field.   
 
In the present work, we use the mixed-phase method near the phase-boundary between the low-field and the $\frac{1}{2}$-plateau phases. Figure \ref{fig:mixed} shows the magnetization curves obtained in the field-increase and field-decrease processes at a fixed temperature. In both the bond-phonon and site-phonon models, a strong hysteresis can clearly be seen, suggestive of a strong first-order transition. We apply the mixed-phase method in this hysteresis region. Following Ref. \cite{MixedMethod_Creutz_79}, we perform relatively short 5000-20000 MC sweeps to obtain the final state and determine the phase boundary between the low-field and middle-field phases shown in the phase diagrams in Figs. \ref{fig:HT_bond} and \ref{fig:HT_siteall}. 

Once the final state is obtained by using the mixed-phase method, we calculate the thermal average of the various physical quantities in the same procedure as that explained in Sec. III: we perform $10^6$ MC sweeps for thermalization and successive $10^6$ MC sweeps for observations. The so-obtained magnetization curves are indicated by the black symbols in Fig. \ref{fig:mixed}. Throughout this paper, we regard the thermal-averaged value obtained in this way as the equilibrium value.

\section{Magnetic structures of the P4 and P5 phases}
\begin{figure}[t]
\begin{center}
\includegraphics[scale=0.6]{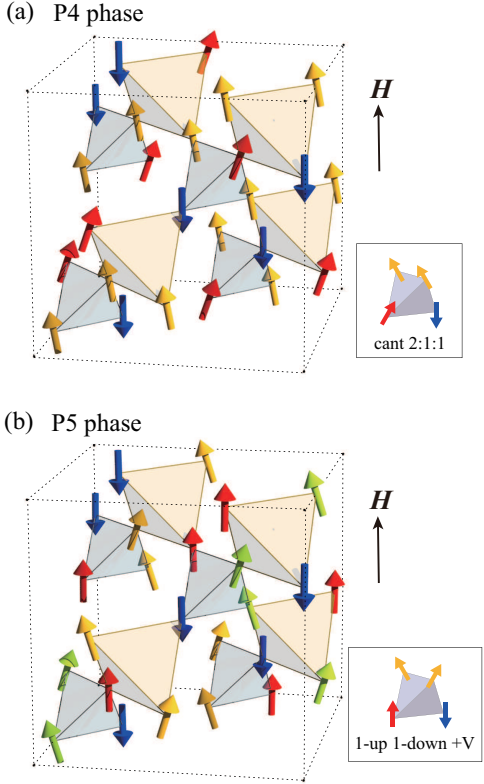}
\caption{Real-space spin structures of (a) the P4 and (b) the P5 phases appearing between the low-field P1 and the middle-field P2 phase. The P4 and P5 phases consist of the ``cant 2:1:1'' and ``1-up, 1-down, and V'' tetrahedra, respectively. Red, yellow, and green arrows represent spins pointing upward along the applied magnetic field $H$, whereas blue ones represent spins pointing downward. The in-plane components of the yellow or green spins at the two corners of a small tetrahedron are parallel in (a), whereas antiparallel in (b). In (b), although in-plane components of the same color spins are collinearly aligned, those of the different color spins, i.e., yellow and green ones, are not necessarily collinear (for details, see the text in Appendix B). \label{fig:cant211U}}
\end{center}
\end{figure}
Figures \ref{fig:cant211U} (a) and (b) show the real-space structures of the P4 and P5 phases, respectively. Local spin configurations on each tetrahedron in the P4 and P5 phases are the cant 2:1:1 state \cite{Bond_Penc_04,Bond_Shannon_10} and the ``1-up, 1-down, and V'' state, respectively. The difference between the two mainly consists in the paired spins at the two corners of a small tetrahedron (see yellow and green arrows in Fig. \ref{fig:cant211U}): the in-plane components of the paired spins are parallel in the P4 phase, whereas antiparallel constituting the ``V'' structure in the P5 phase. 
As mentioned in the main text, in the uniform and weakly breathing cases of $J'/J=1$ and 0.6, the P4 phase is degenerate with the P5 phase at $T=0$, so that both can be the ground state of the site-phonon Hamiltonian (\ref{eq:Hamiltonian_SP}).

In the P5 phase, the in-plane components of the ``V'' spins are collinearly aligned when they are connected by a bond, but not when they are separated by the up or down spin. In the case of Fig. \ref{fig:cant211U} (b), the yellow spins are connected by a bond, and thus, their in-plane components are collinear. Such a situation is also the case for the green spins. Concerning the relative angle between the yellow and green spins separated by the up (red) or down (blue) spins, it can be arbitrary although the relative angle is chosen to be 0 for simplicity in Fig. \ref{fig:cant211U} (b). Such a ground-state degeneracy with respect to the relative angle occurs because of the following reason. Since the inter-tetrahedron interaction in the site-phonon Hamiltonian (\ref{eq:Hamiltonian_SP}) takes the form of $({\bf S}_i\cdot {\bf S}_j)({\bf S}_j\cdot {\bf S}_k)$, the in-plane components of ${\bf S}_i$ and ${\bf S}_k$ do not come into play when ${\bf S}_j$ is aligned along the field direction. Thus, when the ``V'' spins are separated by the up or down spin, their in-plane components perpendicular to the field are not necessarily collinear.

At finite temperatures, it is found by means of MC simulations that among the degenerate P4 and P5 phases, the P5 phase is selected by the thermal fluctuation. This suggests that the entropy in the P5 phase is higher than that in the P4 phase, which is probably because the P5 phase possesses the ground-state degeneracy with respect to the relative angle between the in-plane components of the separated ``V'' spins. In relation to this, the specific heat $C$ becomes lower than 1 in the low-temperature P5 phase, which is in sharp contrast to the conventional behavior typical to classical Heisenberg spin systems, $C \rightarrow 1$ $(T\rightarrow 0)$. Such an unconventional behavior should be relevant to the ground-state degeneracy, but we will leave a further analysis of this issue for our future work.

\section{Magnetic structures of the SP2, SP2', SP3, and SP4 phases}
\begin{figure*}[t]
\begin{center}
\includegraphics[scale=0.52]{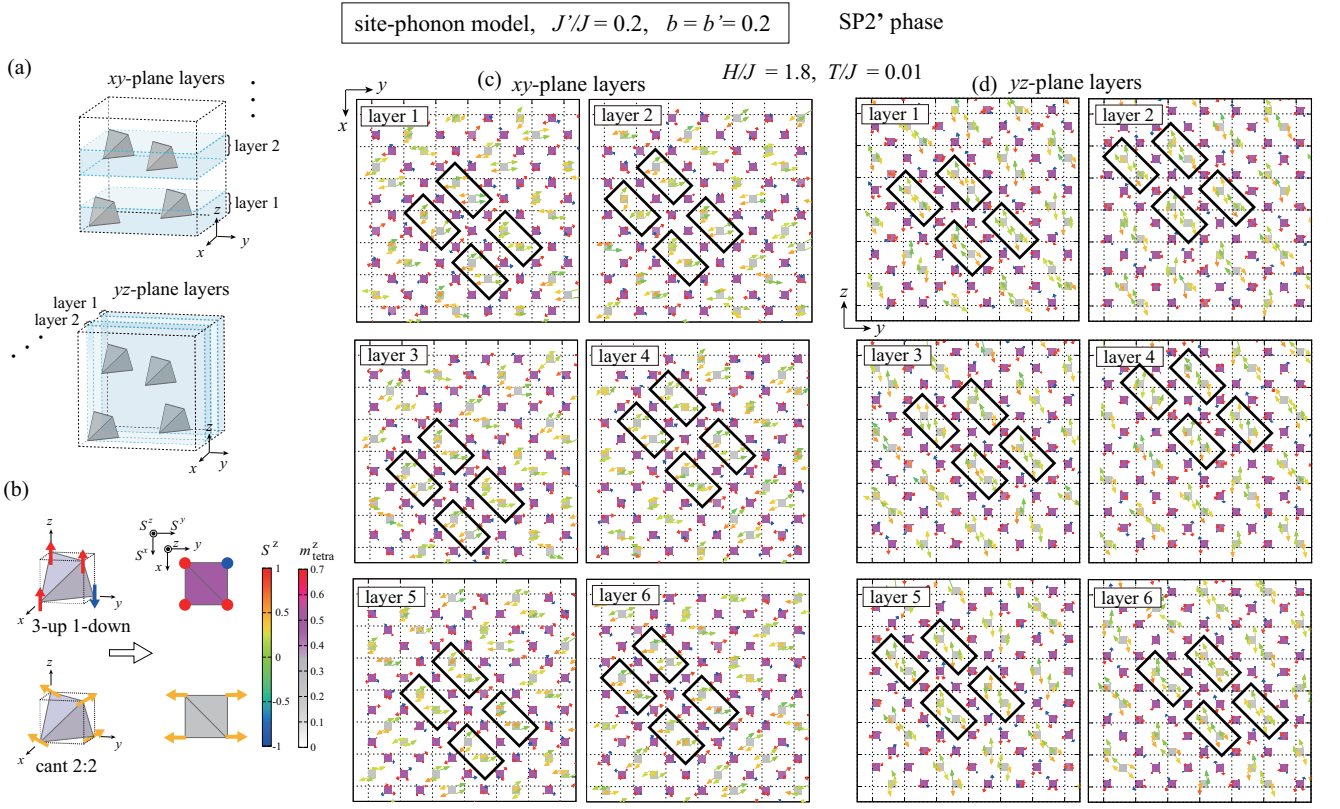}
\caption{Spin structure of the SP2' phase appearing in the site-phonon model with $b=b'=0.20$ in the strongly breathing case of $J'/J=0.2$. (a) The definition of the $xy$-plane (upper) and $yz$-plane (lower) layers of one-small-tetrahedron width. (b) Elementary spin configurations on each tetrahedron (left) and their projections onto the $xy$-plane (right), where in the right, the arrow and its color represent the $S^xS^y$ and $S^z$ components of a spin, respectively, and the color of the box represents the averaged magnetization of each small tetrahedron $m_{\rm tetra}^z$. (c) and (d) MC snapshots taken at $H/J=1.8$ and $T/J=0.01$ in (c) the $xy$-plane layers and (d) the $yz$-plane layers, where the notations are the same as those of (b). A pair of the cant 2:2 tetrahedra is enclosed by a solid black box, and a thin dashed square denotes the cubic unit cell projected onto the two-dimensional plane. \label{fig:snap_site020J02H0180}}
\end{center}
\end{figure*}

\subsection{SP2 and SP2' phases}
As we will explain below, the real-space structures of the higher-temperature SP2 and lower-temperature SP2' phases are essentially the same, so that, here, we will mainly focus on the lower-temperature SP2' phase.
Figures \ref{fig:snap_site020J02H0180} (c) and (d) show the layer-resolved spin configuration of the SP2' phase, where an arrow and its color represent the $S^xS^y$ and $S^z$ components of a spin and a colored square represents a small tetrahedron projected onto the two-dimensional plane. The color of the square denotes the magnetization of each tetrahedron defined by $m_{\rm tetra}^z=(S_1^z+S_2^z+S_3^z+S_4^z)/4$ with four spins at the corners of the tetrahedron ${\bf S}_1$, ${\bf S}_2$, ${\bf S}_3$, and ${\bf S}_4$. The fundamental elements of the SP2' phase are the cant 2:2 tetrahedra and the 3-up and 1-down ones shown in Fig. \ref{fig:snap_site020J02H0180} (b). In the SP2' phase, a pair of the cant 2:2 tetrahedra, whose representative examples are enclosed by thick-line boxes in Figs. \ref{fig:snap_site020J02H0180} (c) and (d), is a building block and the blocks are periodically embedded in the background of the 3-up and 1-down tetrahedra. 
On the layers extending in the $xy$ plane shown in Fig. \ref{fig:snap_site020J02H0180} (c), the cant 2:2 block are stacked right above along the $[\overline{1}10]$ direction, whereas on the $yz$-plane layers shown in Fig. \ref{fig:snap_site020J02H0180} (d), the blocks are stacked zigzag along the $[011]$ direction. The block arrangement pattern on the $zx$-plane layers is the same as the latter, so that it is not shown in Fig. \ref{fig:snap_site020J02H0180}. As one can see from Fig. \ref{fig:snap_site020J02H0180} (c), the $xy$-plane layers are stacked along the $z$-axis, sliding them with their in-plane block patterns being all the same, and such a situation is also the case for the $yz$-plane layers stacking along the $x$-axis [see Fig. \ref{fig:snap_site020J02H0180} (d)]. 

Since on the layers perpendicular to the $x$ and $y$ axes, the same zigzag ordering pattern is realized but not on the layers perpendicular to the $z$ axis, the spin structure should be tetragonal-symmetric with respect to the $z$ axis. 

So far, we have discussed the SP2' phase. Now, we address the real-space structure of the higher-temperature SP2 phase. As mentioned in the main text, the $S^xS^y$ components of spins are ordered in the SP2' phase, while not in the SP2 phase. On the other hand, the spin structure factor for the $S_z$ component $F_{S\parallel}({\bf q})$ in the SP2 phase is almost unchanged from that in the SP2' phase. As $F_{S\parallel}({\bf q})$'s in the SP2 and SP2's phases are almost unchanged, the real-space structure discussed above based on the distributions of $m_{\rm tetra}^z$ is common to these two phases, although the structure is less distinct in the higher-temperature SP2 phase because of the thermal noise. 

\begin{figure*}[t]
\begin{center}
\includegraphics[scale=0.77]{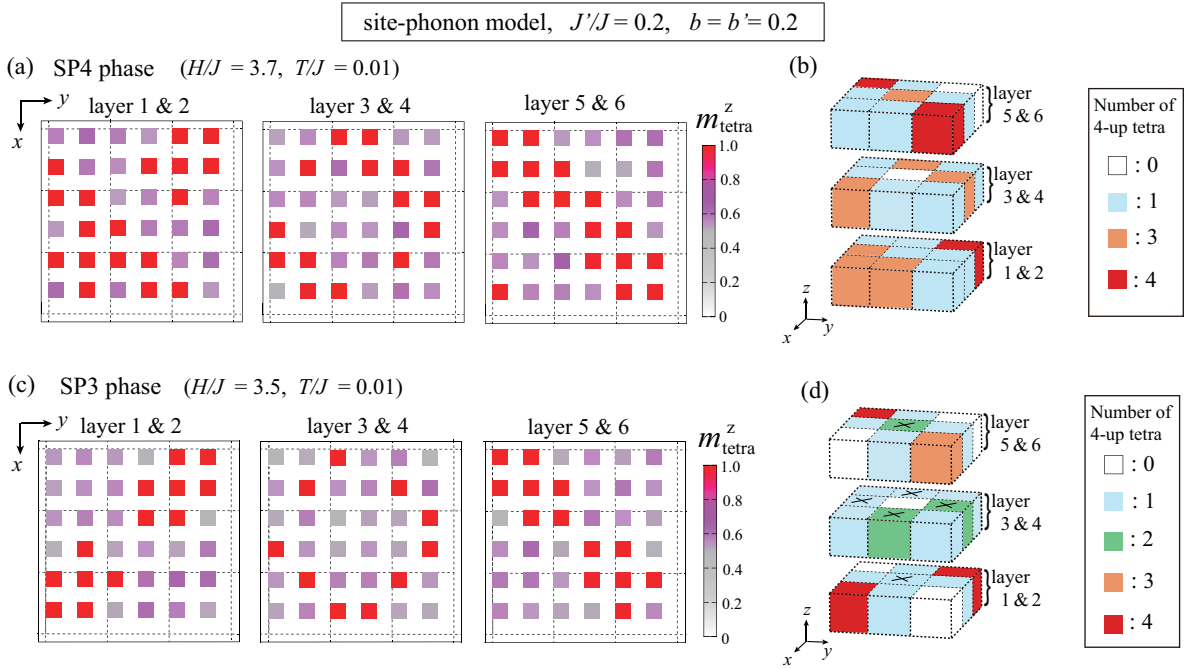}
\caption{MC snapshots of the averaged magnetization of each small tetrahedron $m_{\rm tetra}^z$ obtained in (a) the SP4 phase at $H/J=3.7$ and (c) the SP3 phase at $H/J=3.5$ for the site-phonon model with $b=b'=0.20$ and $J'/J=0.2$, where a small square and its color represent a small tetrahedron and its $m_{\rm tetra}^z$, respectively. In contrast to Figs. \ref{fig:snap_site020J02H0180} (c) and (d), each panel of (a) and (c) represents the distribution of $m_{\rm tetra}^z$ within adjacent two layers defined in Fig. \ref{fig:snap_site020J02H0180} (a), so that the cubic unit cell projected onto the two-dimensional plane, which is indicated by a dotted-line square, contains 4 small tetrahedra. (b) and (d) Schematically-drawn three-dimensional views of the distributions of the cubic unit cells associated with (a) and (c), where red, orange, green, skyblue, and white boxes represent the cubic unit cells containing four, three, two, one, and zero 4-up tetrahedra, respectively. In (d), ordering patterns of the cubes indicated by a cross differ run to run (for details, see the text in Appendix C). 
\label{fig:snap_site020J02H0350-0370}}
\end{center}
\end{figure*}

\subsection{SP3 and SP4 phases}
Figure \ref{fig:snap_site020J02H0350-0370} shows the real-space distributions of $m_{\rm tetra}^z$ in the SP3 and SP4 phases, where a dotted-line box represents a cubic unit cell projected onto the two-dimensional plane. Similarly to the SP1 phase shown in Fig. \ref{fig:snap_site020J06}, the distribution of the 4-up small tetrahedra which correspond to red squares with $m_{\rm tetra}^z=1$ in Fig. \ref{fig:snap_site020J02H0350-0370} is characteristic. In the SP4 phase shown in Fig. \ref{fig:snap_site020J02H0350-0370} (a), there exist four types of cubic unit cells each containing zero, one, three, and four 4-up small tetrahedra, and these cubes are arranged periodically. The three-dimensional view of the cube arrangement is schematically shown in Fig. \ref{fig:snap_site020J02H0350-0370} (b). The state is 3-fold symmetric with respect to the $[\overline{1}11]$ direction, but not for other related directions such as $[111]$, so that the SP4 phase is non-cubic. In the same manner, one can classify the ordering pattern in the SP3 phase. As shown in Figs. \ref{fig:snap_site020J02H0350-0370} (c) and (d), the SP3 phase consists of all the five types of the cubic unit cells, and they are built up in such a way that some of them constitute the basis structure of the ordering pattern and others a non-basis part whose ordering pattern is not uniquely determined. In Fig. \ref{fig:snap_site020J02H0350-0370} (d), the former (latter) correspond to boxes without (with) a cross. Since in Fig. \ref{fig:snap_site020J02H0350-0370} (d), the basis structure is 3-fold symmetric with respect to the $[111]$ direction but not for other related directions, the state is non-cubic. Concerning the non-basis part, two types of the ordering patterns are obtained in the MC simulations: in one case, a cubic unit cell containing two 4-up tetrahedra [a green box in Fig. \ref{fig:snap_site020J02H0350-0370} (d)] is located in the layer with the cubic unit cell containing three 4-up tetrahedra [the top layer containing an orange box in Fig. \ref{fig:snap_site020J02H0350-0370} (d)], whereas in the other case, it is located in the layer without the one [the bottom layer without an orange box in Fig. \ref{fig:snap_site020J02H0350-0370} (d)]. In both cases, the basis structure remains unchanged. We note that these results are obtained for $L=3$ which is the smallest size for the SP3 phases. Cube arrangement patterns for the larger sizes of $L=6$ and 12 look rather complicated but they follow the above ordering rule part by part, which is probably due to the existence of the two types of the non-basis ordering patterns. By contrast, in the SP4 phase, the cube arrangement pattern presented in Figs. \ref{fig:snap_site020J02H0350-0370} (a) and (b) is uniquely determined, so that its periodic structure can be obtained in the larger-size systems.


\begin{thebibliography}{100}
\bibitem{ZnCrO_Lee_00} S.-H. Lee, C. Broholm, T. H. Kim, W. Ratcliff II, and S-W. Cheong, Phys. Rev. Lett. {\bf 84}, 3718 (2000).
\bibitem{CdCrO_Chung_05} J.-H. Chung, M. Matsuda, S.-H. Lee, K. Kakurai, H. Ueda, T. J. Sato, H. Takagi, K.-P. Hong, and S. Park, Phys. Rev. Lett. {\bf 95}, 247204 (2005).
\bibitem{HgCrO_Ueda_06} H. Ueda, H. Mitamura, T. Goto, and Y. Ueda, Phys. Rev. B {\bf 73}, 094415 (2006).
\bibitem{MgCrO_Ortega_08} L. Ortega-San-Martin, A. J. Williams, C. D. Gordon, S. Klemme, and J. P. Attfield, J. Phys.: Condens. Matter {\bf 20}, 104238, (2008).

\bibitem{CdCrO_Kojima_08} E. Kojima, A. Miyata, S. Miyabe, S. Takeyama, H. Ueda, and Y. Ueda, Phys. Rev. B {\bf 77}, 212408 (2008).
\bibitem{CdCrO_Miyata_13} A. Miyata, S. Takeyama and H. Ueda, Phys. Rev. B {\bf 87}, 214424 (2013). 
\bibitem{ZnCrO_Miyata_jpsj_11} A. Miyata, H. Ueda, Y. Ueda, Y. Motome, N. Shannon, K. Penc, and S. Takeyama, J. Phys. Soc. Jpn. {\bf 80}, 074709 (2011).
\bibitem{ZnCrO_Miyata_prl_11} A. Miyata, H. Ueda, Y. Ueda, H. Sawabe, and S. Takeyama, Phys. Rev. Lett. {\bf 107}, 207203 (2011).
\bibitem{ZnCrO_Miyata_jpsj_12} A. Miyata, H. Ueda, Y. Ueda, Y. Motome, N. Shannon, K. Penc, and S. Takeyama, J. Phys. Soc. Jpn. {\bf 81}, 114701 (2012).
\bibitem{HgCrO_Nakamura_jpsj_14} D. Nakamura, A. Miyata, Y. Aida, H. Ueda, and S. Takeyama, J. Phys. Soc. Jpn. {\bf 83}, 113703 (2014). 
\bibitem{MgCrO_Miyata_jpsj_14} A. Miyata, H. Ueda, and S. Takeyama, J. Phys. Soc. Jpn. {\bf 83}, 063702 (2014).

\bibitem{BrPyro_Okamoto_13} Y. Okamoto, G. J. Nilsen, J. P. Attfield, and Z. Hiroi, Phys. Rev. Lett. {\bf 110}, 097203 (2013).
\bibitem{BrPyro_Tanaka_14} Y. Tanaka, M. Yoshida, M. Takigawa, Y. Okamoto, and Z. Hiroi, Phys. Rev. Lett. {\bf 113}, 227204 (2014).
\bibitem{BrPyro_Nilsen_15} G. J. Nilsen, Y. Okamoto, T. Masuda, J. Rodriguez-Carvajal, H. Mutka, T. Hansen, and Z. Hiroi, Phys. Rev. B {\bf 91}, 174435 (2015).
\bibitem{BrPyro_Saha_16} R. Saha, F. Fauth, M. Avdeev, P. Kayser, B. J. Kennedy, and A. Sundaresan, Phys. Rev. B {\bf 94}, 064420 (2016).
\bibitem{BrPyro_Lee_16} S. Lee, S.-H. Do, W.-J. Lee, Y. S. Choi, M. Lee, E. S. Choi, A. P. Reyes, P. L. Kuhns, A. Ozarowski, and K.-Y. Choi, Phys. Rev. B {\bf 93}, 174402 (2016).
\bibitem{BrPyro_Hdep_Okamoto_17} Y. Okamoto, D. Nakamura, A. Miyake, S. Takeyama, M. Tokunaga, A. Matsuo, K. Kindo, and Z. Hiroi, Phys. Rev. B {\bf 95}, 134438 (2017).
\bibitem{BrPyro_Hdep_Gen_19} M. Gen, D. Nakamura, Y. Okamoto, and S. Takeyama, J. Magn. Magn. Mater. {\bf 473}, 387 (2019).

\bibitem{Bond_Penc_04} K. Penc, N. Shannon, and H. Shiba, Phys. Rev. Lett. {\bf 93}, 197203 (2004).
\bibitem{Bond_Motome_06} Y. Motome, K. Penc, and N. Shannon, J. Magn. Magn. Mater. {\bf 300}, 57 (2006).
\bibitem{Bond_Shannon_10} N. Shannon, K. Penc, and Y. Motome, Phys. Rev. B {\bf 81}, 184409 (2010).
\bibitem{Site_Jia_05} C. Jia, J. H. Nam, J. S. Kim, and J. H. Han, Phys. Rev. B {\bf 71}, 212406 (2005). 
\bibitem{Site_Bergman_06} D. L. Bergman, R. Shindou, G. A. Fiete, and L. Balents, Phys. Rev. B {\bf 74}, 134409 (2006).
\bibitem{Site_Wang_08} F. Wang and A. Vishwanath, Phys. Rev. Lett. {\bf 100}, 077201 (2008). 
\bibitem{Site_AK_16} K. Aoyama and H. Kawamura, Phys. Rev. Lett. {\bf 116}, 257201 (2016).  
\bibitem{Site_AK_19} K. Aoyama and H. Kawamura, Phys. Rev. B {\bf 99}, 144406 (2019).  

\bibitem{Reimers_MC_92} J. N. Reimers, Phys. Rev. B {\bf 45}, 7287 (1992).
\bibitem{Moessner-Chalker_prl} R. Moessner and J. T. Chalker, Phys. Rev. Lett. {\bf 80}, 2929 (1998).
\bibitem{Moessner-Chalker_prb} R. Moessner and J. T. Chalker, Phys. Rev. B {\bf 58}, 12049 (1998).

\bibitem{HgCrO_Matsuda_07} M. Matsuda, H. Ueda, A. Kikkawa, Y. Tanaka, K. Katsumata, Y. Narumi, T. Inami, Y. Ueda, and S.-H. Lee, Nat. Phys. {\bf 3}, 397 (2007).
\bibitem{CdCrO_Inami_06} T. Inami, K. Ohwada, M. Tsubota, Y. Murata, Y. H. Matsuda, H. Nojiri, H. Ueda, and Y. Murakami, J. Phys.: Conf. Ser. {\bf 51}, 502-505, (2006). 
\bibitem{CdCrO_Matsuda_prl_10} M. Matsuda, K. Ohoyama, S. Yoshii, H. Nojiri, P. Frings, F. Duc, B. Vignolle, G. L. J. A. Rikken, L.-P. Regnault, S.-H. Lee, H. Ueda, and Y. Ueda, Phys. Rev. Lett. {\bf 104}, 047201 (2010).

\bibitem{BrPyro_Saha_17} R. Saha, R. Dhanya, C. Bellin, K. B\'{e}neut, A. Bhattacharyya, A. Shukla, C. Narayana, E. Suard, J. Rodr\'{i}guez-Carvajal, and A. Sundaresan, Phys. Rev. B {\bf 96}, 214439 (2017).  
\bibitem{BrPyro_doped_Okamoto_15} Y. Okamoto, G. J. Nilsen, T. Nakazono, and Z. Hiroi, J. Phys. Soc. Jpn. {\bf 84}, 043707 (2015).
\bibitem{BrPyro_doped_Wang_17} D. Wang, C. Tan, K. Huang, and L. Shu, Chinese Phys. Lett. {\bf 33}, 127501 (2016).  
\bibitem{BrPyro_doped_Wawrzynczak_17} R. Wawrzy\'{n}czak, Y. Tanaka, M. Yoshida, Y. Okamoto, P. Manuel, N. Casati, Z. Hiroi, M. Takigawa, and G. J. Nilsen, Phys. Rev. Lett. {\bf 119}, 087201 (2017).
\bibitem{BrPyro_Sulfides_Okamoto_18} Y. Okamoto, M. Mori, N. Katayama, A. Miyake, M. Tokunaga, A. Matsuo, K. Kindo, and K. Takenaka, J. Phys. Soc. Jpn. {\bf 87}, 034709 (2018).
\bibitem{BrPyro_Sulfides_Pokharel_18} G. Pokharel, A. F. May, D. S. Parker, S. Calder, G. Ehlers, A. Huq, S. A. J. Kimber, H. S. Arachchige, L. Poudel, M. A. McGuire, D. Mandrus, and A. D. Christianson, Phys. Rev. B {\bf 97}, 134117 (2018).
\bibitem{BrPyro_Hdep_Gen_20} M. Gen, Y. Okamoto, M. Mori, K. Takenaka, and Y. Kohama, Phys. Rev. B {\bf 101}, 054434 (2020).
\bibitem{BrPyro_Sulfides_Kanematsu_20} T. Kanematsu, M. Mori, Y. Okamoto, T. Yajima, and K. Takenaka, J. Phys. Soc. Jpn. {\bf 89}, 073708 (2020).
\bibitem{BrPyro_Sulfides_Pokharel_20} G. Pokharel, H. S. Arachchige, T. J. Williams, A. F. May, R. S. Fishman, G. Sala, S. Calder, G. Ehlers, D. S. Parker, T. Hong, A. Wildes, D. Mandrus, J. A. M. Paddison, and A. D. Christianson, Phys. Rev. Lett. {\bf 125}, 167201 (2020).
\bibitem{qBrPyro_Kimura_14} K. Kimura, S. Nakatsuji, and T. Kimura, Phys. Rev. B {\bf 90}, 060414(R) (2014).  
\bibitem{qBrPyro_Haku_prb16} T. Haku, K. Kimura, Y. Matsumoto, M. Soda, M. Sera, D. Yu, R. A. Mole, T. Takeuchi, S. Nakatsuji, Y. Kono, T. Sakakibara, L.-J. Chang, and T. Masuda, Phys. Rev. B {\bf 93}, 220407(R) (2016).  
\bibitem{qBrPyro_Haku_jpsj16} T. Haku, M. Soda, M. Sera, K. Kimura, S. Itoh, T. Yokoo, and T. Masuda, J. Phys. Soc. Jpn. {\bf 85}, 034721 (2016).  
\bibitem{qBrPyro_Rau_16} J. G. Rau, L. S. Wu, A. F. May, L. Poudel, B. Winn, V. O. Garlea, A. Huq, P. Whitfield, A. E. Taylor, M. D. Lumsden, M. J. P. Gingras, and A. D. Christianson, Phys. Rev. Lett. {\bf 116}, 257204 (2016).
\bibitem{qBrPyro_Rau_18} J. G. Rau, L. S. Wu, A. F. May, A. E. Taylor, I-Lin. Liu, J. Higgins, N. P. Butch, K. A. Ross, H. S. Nair, M. D. Lumsden, M. J. P. Gingras, and A. D. Christianson, J. Phys.: Condens. Matter {\bf 30}, 455801 (2018).


\bibitem{BrPyro_NNmodel_Benton_15} O. Benton and N. Shannon, J. Phys. Soc. Jpn. {\bf 84}, 104710 (2015).

\bibitem{SLC_Yamashita_00} Y. Yamashita and K. Ueda, Phys. Rev. Lett. {\bf 85}, 4960 (2000).
\bibitem{SLC_Tchernyshyov_prl_02} O. Tchernyshyov, R. Moessner, and S. L. Sondhi, Phys. Rev. Lett. {\bf 88}, 067203 (2002).
\bibitem{SLC_Tchernyshyov_prb_02} O. Tchernyshyov, R. Moessner, and S. L. Sondhi, Phys. Rev. B {\bf 66}, 064403 (2002).

\bibitem{ZnCrO_Lee_08} S.-H. Lee, W. Ratcliff, Q. Huang, T. H. Kim, and S-W. Cheong, Phys. Rev. B {\bf 77}, 014405 (2008).
\bibitem{SLC_Chern_06} Gia-Wei Chern, C. J. Fennie, and O. Tchernyshyov, Phys. Rev. B {\bf 74}, 060405(R) (2006).
\bibitem{HgCrO_Kimura_jpsj_14} S. Kimura, M. Hagiwara, T. Takeuchi, H. Yamaguchi, H. Ueda, and K. Kindo, J. Phys. Soc. Jpn. {\bf 83}, 113709 (2014).
\bibitem{CdCrO_Rossi_prl_19} L. Rossi, A. Bobel, S. Wiedmann, R. K\"{u}chler, Y. Motome, K. Penc, N. Shannon, H. Ueda, and B. Bryant, Phys. Rev. Lett. {\bf 123}, 027205 (2019), Supplemental Material. 

\bibitem{Loop_Shinaoka_14} H. Shinaoka, Y. Tomita, and Y. Motome, Phys. Rev. B {\bf 90}, 165119 (2014).
\bibitem{Fukushima_exchange} K. Hukushima and K. Nemoto, J. Phys. Soc. Jpn. {\bf 65}, 1604 (1996).
\bibitem{MixedMethod_Creutz_79} M. Creutz, L. Jacobs, and C. Rebbi, Phys. Rev. D {\bf 20}, 1915 (1979).   

\bibitem{CdCrO_Kimura_jpsj_15} S. Kimura, Y. Sawada, Y. Narumi, K. Watanabe, M. Hagiwara, K. Kindo, and H. Ueda, Phys. Rev. B {\bf 92}, 144410 (2015). 
\bibitem{FirstPrinciple_Ghosh_npj_19} P. Ghosh, Y. Iqbal, T. Müller, R. T. Ponnaganti, R. Thomale, R. Narayanan, J. Reuther, M. J. P. Gingras, and H. O. Jeschke, npj Quantum Mater. {\bf 4}, 63 (2019).

\end{thebibliography}
\end{document}